\documentstyle[epsf]{mn}

\title[Field Ellipticals at $z\approx 0.3$]{The Properties of Field
Elliptical
Galaxies at Intermediate Redshift.  I: Empirical Scaling Laws
\footnote{Based on observations collected at the European Southern
Observatory, La Silla and with the NASA/ESA Hubble Space Telescope,
obtained at the Space Telescope Science Institute, which is operated
by Association of Universities for Research in Astronomy, Inc.\
(AURA), under NASA contract NAS5-26555.}}

\author[T.~Treu et al.]{T.~Treu,$^{1,2,6}$ M.~Stiavelli,$^{2,3,4}$
S.~Casertano,$^{2,3}$ P.~M{\o}ller,$^{2,3,5}$ G.~Bertin$^1$\\
$^1$ Scuola Normale Superiore, P.za dei Cavalieri 7, I56126, Pisa,
Italy\\
$^2$ Space Telescope Science Institute, 3700 San Martin Dr.,
Baltimore, MD 21218, USA\\
$^3$ On assignment from the Space Sciences Division of the European
Space Agency\\
$^4$ On leave from the Scuola Normale Superiore, Pisa, Italy\\
$^5$ ESO, Karl-Schwarzschild Str. 2, D85748, Garching bei M\"unchen,
Germany\\
$^6$ e-mail address: treu@cibs.sns.it
}

\newcommand{\tx}[1]{\textrm{#1}}
\newcommand{\pc}[1]{\protect\citename{#1}}
\newcommand{\sfq}{$\sigma_{\tx{fq}}$\,}
\newcommand{\kms}{km~$\tx{s}^{-1}$}
\newcommand{\gfq}{$\gamma_{\tx{fq}}$\,}
\newcommand{\dv}{$r^{1/4}\,$}
\newcommand{\sgh}{$\sigma_{\tx{ghff}}$\,}
\newcommand{\ggh}{$\gamma_{\tx{ghff}}$\,}
\newcommand{\zgh}{$z_{\tx{ghff}}$\,}
\newcommand{\zfq}{$z_{\tx{fq}}$\,}
\begin{document}
\maketitle
\begin{abstract}

We present measurements of the Fundamental Plane (FP)
parameters (the effective radius, the mean effective surface
brightness, and the central velocity dispersion) of six field
elliptical galaxies at intermediate redshift.  The imaging is taken
from the Medium Deep Survey of the Hubble Space Telescope, while the
kinematical data are obtained from long-slit spectroscopy using the
3.6-m ESO telescope.  The Fundamental Plane appears well defined in
the field even at redshift $\approx$ 0.3. The data show a shift in the
FP zero point with respect to the local relation, possibly indicating
modest evolution, consistent with the result found for intermediate
redshift cluster samples.  The FP slopes derived for our field data,
plus other cluster ellipticals at intermediate redshift taken from
the literature, differ from the local ones, but are still consistent
with the interpretation of the FP as a result of homology, of the
virial theorem and of the existence of a relation between luminosity
and mass, $L \propto M^{\eta}$. We also derive the surface brightness
vs. effective radius relation for nine galaxies with redshift up to $z
\approx0.6$, and data from the literature; the evolution that can be
inferred is consistent with what is found using the FP.
\end{abstract}

\section{Introduction}
\label{sec:intro}
Elliptical galaxies in the local universe are known to populate only a
two-dimensional manifold, known as the Fundamental Plane (hereafter
FP; Djorgovski \& Davis 1987; Dressler et al.~1987), in the
three-dimensional space defined by effective radius $R_{\tx{e}}$, mean
surface brightness within the effective radius SB$_{\tx{e}}$
(SB$_{\tx{e}}=-2.5\log\langle I \rangle_{\tx{e}}+$const; I is the
surface brightness in linear flux), and central velocity dispersion~$\sigma$. The FP is described by:
\begin{equation}
\label{eq:FP} \log R_{\tx{e}} = \alpha \log~\sigma + \beta~\tx{SB}_{\tx{e}} + \gamma,  
\end{equation} 
$R_{\tx{e}}$ is in kpc (while $r_{\tx{e}}$ is in arcsecs), $\sigma$ in
\kms, SB$_{\tx{e}}$ in $\tx{mag arcsec}^{-2}$ and $H_0$ is hereafter
assumed to be 50 $h_{50}$ \kms $\tx{Mpc}^{-1}$. Typical values of
the coefficients are for example $\alpha=1.25$, $\beta=0.32$ and
$\gamma=-8.895$ in Johnson B (Bender et al.\ 1998; hereafter B98).  It is still
unclear whether the thickness of the FP is intrinsic, i.e., due to
scatter in the properties of elliptical galaxies, or is due to
observational errors.

The existence of the empirical scaling described by the FP has strong
implications in terms of galactic evolution and formation theories.
For example, the FP suggests that the distributions of dark and
luminous matter are related, since $\sigma$ depends on the total
gravitational potential, whereas SB$_{\tx{e}}$ and $R_{\tx{e}}$ trace
only the stellar mass. The FP can be explained in terms of homology,
of the virial theorem, and of the existence of a well-defined relation
between luminosity and mass, $ L \propto M^{\eta}$ (Faber et al.,
1987; see also van Albada, Bertin \& Stiavelli~1995). This
interpretation requires a relation between the derived values of
$\alpha$ and $\beta$, namely $\alpha-10\beta+2=0$
($\eta=0.2~\alpha/\beta$).

Measuring the FP parameters for ellipticals at intermediate redshift $
z \approx 0.3$--$0.8 $ has only recently become feasible (Pahre,
Djorgovski \& de Carvalho 1995; van Dokkum \& Franx 1996; Bender et
al.~1996; Kelson et al. 1997; van Dokkum et al. 1998).  This opens up
two important questions which bear on the age, formation history, and
internal properties of elliptical galaxies: {\it i)} how far in the
past does an FP-like relation apply? and {\it ii)} do its parameters
evolve significantly with time? The existence of a tight FP-like
relation at substantial look-back times would suggest the existence of
a universal relation between mass and formation time for ellipticals,
since, e.g., population differences would be amplified as the redshift
increases. Existing measurements of the properties of ellipticals at
intermediate redshifts have concentrated so far on relatively rich
clusters of galaxies, where bright ellipticals are easier to find
\cite{PDdC95,DF96,B96,ZB97,KDFIF,DFKI98} and observations can be
carried out efficiently with multi-object spectrographs. When
comparing, e.g., the FP of intermediate redshift and zero redshift
clusters one must take into account the evolution with redshift of the
population of galaxies in clusters, and environmental effects.
Although the evolution of clusters of galaxies is still somewhat
controversial \cite{Post96}, the galaxy population in rich clusters is
likely to become less uniform (in terms, e.g., of age and metallicity)
with time, due to accretion of isolated galaxies and small groups.
Moreover, the question of whether there are systematic differences
between cluster and field ellipticals (the latter including galaxies
in loose groups or poor clusters) is still open even at low redshift
\cite{dCD92}.  A study of the galaxy properties as a function of
look-back time provides a sensitive probe of the possible evolutionary
differences between cluster and field ellipticals.  For these reasons,
we have begun a study of the empirical scaling laws of field
ellipticals at intermediate redshift.

We have started out by defining a sample of ellipticals which is not
biased in favour of rich clusters. Details on the selection process
are given in Section~\ref{sec:sample}. Sections~\ref{sec:foto} and
\ref{sec:spec} describe the data reduction and the error analysis.
Measurements at non-zero redshift have to be carried out with
photometric filters and effective apertures that differ from those
used to derive the local scaling laws; Section~\ref{sec:corr}
describes the conversion of our measurements to the standard
quantities.  The main results are reported in Section~\ref{sec:res},
where our measurements are compared to local samples and intermediate
redshift cluster samples. A summary is given in Section~\ref{sec:conc}.

\section{Sample selection}
\label{sec:sample}

With this study we aim at addressing two questions: {\it i)} does the FP of field ellipticals evolve with redshift? and {\it ii)} do, at any
given redshift, the field ellipticals have FP parameters different from those
found in rich cluster environments? For this we need to select a
primary sample of galaxies biased against rich cluster membership
at intermediate redshift, where the project is feasible and the look-back time
($\approx 4$--$5$ Gyrs) is high enough that evolution can be noticeable.

The targets used in this study have been chosen among a sample of
random ellipticals found in the WFPC2 parallel images collected by
the Medium Deep Survey \cite{MDS}.  The target pool has
been selected originally from available HST images in the
appropriate RA, $\delta $ range according to the following criteria:

\begin {enumerate}

\item Apparent magnitude $ I < 20.5 $ 

\item Morphology clearly defined as elliptical, with an apparent
effective radius $ r_{\tx{e}} > 0\farcs 5 $.

\item $ V - I $ colour in agreement with the empiric relation for
ellipticals in the redshift range $ 0.2 \leq z \leq 0.5 $ ($ 1.1 < V-I
< 1.7 $)

\end {enumerate}

Magnitudes and colours above are as defined by the Medium Deep Survey
group \cite{MDS}; $ V $ and $ I $ are the magnitudes in the WFPC2
filters F606W and F814W, computed in the WFPC2 Flight system (Holtzman
et al.~1995), and the effective radius $ r_{\tx{e}} $ is the value found from
the two-dimensional image fitting carried out by the MDS.  All
photometric parameters have been rederived here with two different
techniques (see Section~\ref{sec:foto}) to ensure a
self-consistent treatment.

Individual targets were then selected from this pool of approximately 25
candidates on the basis of convenience and availability during each
observing night; in the spirit of this exploratory study, we did not
attempt further to achieve uniformity or completeness within our pool of
candidates. 

Since we did not explicitly exclude members of poor clusters (no rich
clusters were observed in these random fields), the ellipticals in our
sample should be representative of a random, magnitude-selected sample
of non-rich-cluster objects.  In fact, four of the six objects for
which we have obtained FP parameters happen to be members of a group
(or a poor cluster), which was randomly observed in the available HST
images.  Future observations will enable us to achieve a better level
of completeness and thus to better characterise our sample.

Our colour selection criterion will bias against actively star-forming
ellipticals, which are known to occur, albeit infrequently, in
complete samples \cite{CFRS}.  We plan to relax the blue cutoff in
future observations.

\subsection{Target list}

The characteristics of the objects we observed are summarised in
Table~\ref{tab:obj}. The final sample used in the discussion in
Section~\ref{sec:res} is composed of the galaxies successfully fitted
with the two different photometric techniques, isophotal profile fit and
two-dimensional fit, and with measured kinematics.

The photometry of {\bf C, L, M, N} was of low quality due to
companion galaxies, stars in the field of view and the faintness of the
objects. The 2D fit was performed only on the galaxies with a high
quality isophotal profile.  We were able to determine high quality
photometric parameters for {\bf G} by subtracting a model of its
companion {\bf H} from the image before the 2D fit.

\begin{table}
\caption{Galaxy identification.  (1) Adopted names, (2) HST Medium
Deep Survey field name, (3) WFPC2 chip, (4) ID within the field, (5)
and (6) pixel positions of the center of the galaxy, (7) Sizes of the
F606W and F814W images used in the 2D fits and displayed in Figures 2
and 3. The image used for {\bf I} in F814W was $5\farcs 4 \times 4
\farcs 8$. Dashes in Column 7 indicate that the 2D fit was not
performed.}
\label{tab:obj}
\begin{tabular}{c c c c c c c}
galaxy & field  & chip & n & x   & y & size \\	
 (1)	& (2)	& (3)	& (4)	& (5)	& (6) & (7) \\	
\hline
 A &    ut800 &  4 &    1    & 324 &  444 & $18''\times 18''$ \\
 B &    ut800 &  4 &    2    & 265 &  741 & $8''\times 8''$ \\
 C &    ur610 &  3 &    1    & 678 &  691 & -\\
 D &    ur610 &  3 &    2    & 723 &  642 & $9\farcs 5\times 6\farcs 4$\\
 E &    u5405 &  2 &    1    & 335 &  324 & $8''\times 7''$\\
 F &    u5405 &  2 &    2    & 332 &  221 & $5\farcs 4 \times  6 \farcs 4$\\
 G &    u5405 &  2 &    3    & 192 &  288 & $10'' \times 10''$\\
 H &    u5405 &  2 &    4    & 194 &  255 & -\\
 I &    u5405 &  2 &    5    & 72  &  63  & $5\farcs 2 \times 4 \farcs 3$ \\
 L &    u5405 &  3 &    6    & 254 &  711 & -\\
 M &    u5405 &  3 &    7    & 266 &  715 & -\\
 N &    u5405 &  3 &    8    & 747 &  732 & -\\
 O &    ust00 &  3 &    1    & 226 &  523 & $5'' \times 5''$\\
 P &    ust00 &  3 &    2    & 336 &  404 & $5'' \times 5''$\\
 Q &    ust00 &  3 &    3    & 443 &  298 & $5 \farcs 2 \times 5 \farcs 4$\\
 R &    urz00 &  2 &    1    & 719 &  242 & -\\
\hline
\end{tabular}
\end{table} 

\section{Photometry}
\label{sec:foto}

The images are taken from the Medium Deep Survey of the
Hubble Space Telescope \cite{MDS}. For each field there are images
from WFPC2 through filters F606W and F814W. Table~\ref{tab:phlog}
lists total exposure times and number of exposures.

\begin{table}
\caption{Photometric data: number of exposures (nexp) and total exposure times (texp).}
\label{tab:phlog}
\begin{tabular}{|l|l|r|r|}
field 	& filter 	& nexp 		&  texp	(s)\\
\hline
u5405 	& F606W	&  1 &   800 	\\
	& F814W	&  1 &   800 	\\
ur610	& F606W	&  2 &  1500 	\\
	& F814W	&  2 &  1600 	\\
urz00	& F606W	&  3 &  5400 	\\
	& F814W	&  4 &  8400 	\\
ust00	& F606W	& 10 & 16500 	\\
	& F814W	& 11 & 23100 	\\
ut800	& F606W	&  2 &  1430 	\\
	& F814W	&  3 &  4430 	\\
\hline
\end{tabular} 
\end{table}

\subsection{Reduction}

The basic reduction was done at the Space Telescope Science Institute
using the standard pipeline with the best reference files available.  As
bias and dark files we used the superbias and the superdark from the
Hubble Deep Field \cite{HDF} for their high signal-to-noise ratio.  The
cosmic ray removal for the fields with multiple images was done using
the {\sc iraf/stsdas} task {\sc crrej}, with a cutoff at 5 standard deviations
from the mean.  Since
only one exposure per filter was available for u5405, the original image
of this field was compared with a smoothed version (using the {\sc
midas} command {\sc filter/cosmic}).  In all cases we checked by eye
that the pixels removed in the area of the targets were only those
affected by cosmic rays. 

For u5405, chip 2 of WFPC2 is affected by a bad column, very close to
the center of galaxies {\bf E} and {\bf F} (see Figure~\ref{fig:2D}).
We interpolated through it in the outer parts of the galaxies, but we
chose not to modify the 4-5 pixel diameter central zone because it was
too steep for any interpolation to be meaningful (see also
Figure~\ref{fig:profs}).  In order to test how much this bad column
affected the determination of the photometric parameters, we fitted
the isophotal profile of {\bf E} excluding the innermost part of the
profile. We tried excluding a 2 pixel and a 4 pixel diameter circle;
the effective radius and the effective surface brightness change
significantly (up to 20\% in $r_{\tx{e}}$), towards the value given by
the 2D fit (smaller $r_{\tx{e}}$ and brighter surface
brightness). These new values are still consistent with the average
value within the internal errors (that are the largest of the sample, 
see Subsection~\ref{ssec:fotoerr}). Furthermore it should be noticed
that the derived values for $r_{\tx{e}}$ and SB$_{\tx{e}}$ are highly
correlated and therefore the variation of the combination that enters
the FP, $\log r_{\tx{e}} - \beta \tx{SB}_{\tx{e}}$, is much smaller
than the variation on the single coefficients (see
Subsection~\ref{ssec:fotoerr} and Table~\ref{tab:fotcorr}).

A constant sky contribution was subtracted from each image.  The sky
contribution used was the mean of the averages of areas with no evidence
for light sources.  This procedure does not take into account smooth and
diffuse background sources, which is one of the main sources of uncertainty
when using curve of growth technique (see Section~\ref{ssec:fotoerr}). 
  
\subsection{Isophotes and two-dimensional fits}

The photometric parameters needed for the FP are the effective radius
$r_{\tx{e}}$, and the effective surface brightness SB$_{\tx{e}}$,
measured as the best \dv light profiles parameters.

In order to obtain the best results and robust error estimates, the
photometric parameters were derived using two independent techniques:
fits to the isophotal luminosity profiles and two-dimensional fits to
the images.  Both procedures require a PSF to convolve the models
with.  We used synthetic PSFs calculated with Tiny Tim 4.0 \cite{TT4}
with a 15 mas jitter.  The quality of the synthetic PSF was checked by
comparing the isophotal profiles of two stars and the profiles of two
PSFs created in the same spot.  No significant difference was
noticed. As a double-check we fitted an isophotal profile using both a
synthetic PSF profile and a real star profile: the differences were
less than 1\% in $r_{\tx{e}}$ and less than 0.02 mag in
SB$_{\tx{e}}$. The photometric parameters of each galaxy were derived
independently in the two bandpasses F814W and F606W.

\subsubsection{Isophotes}
\label{sssec:iso}

\begin{figure*}
\mbox{\epsfysize=18cm \epsfbox{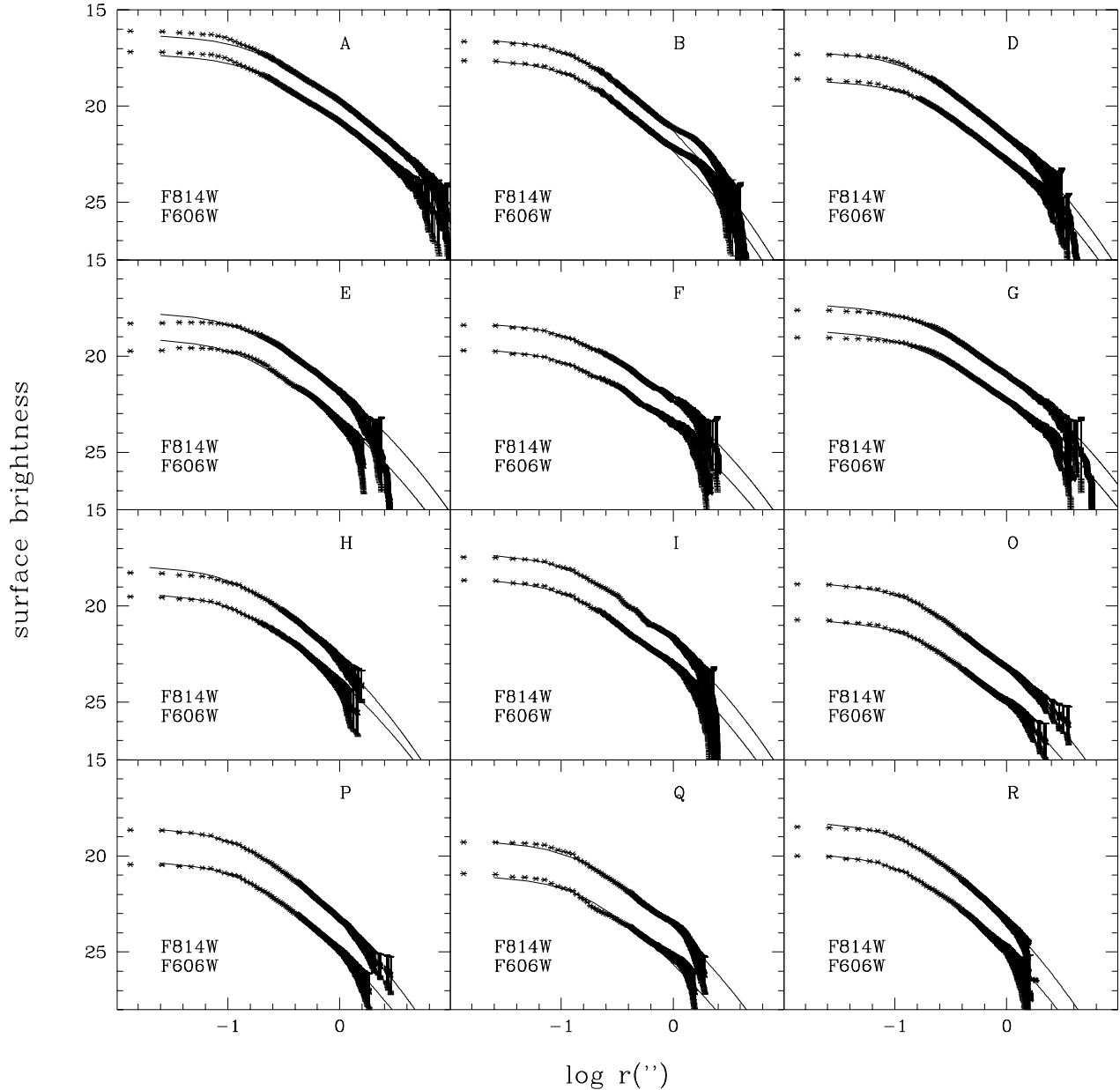}}
\caption{Isophotal profiles.  Crosses are the data points, solid lines
represent the best-fit \dv profile, convolved with the PSF.  The upper
data refer to F814W, the lower to F606W.  The error bars (which are
smaller than the crosses for the innermost part of the profile) take
into account the sky-subtraction error. The bump on the {\bf B}
profile is possibly the signature of a disk component (see text,
Section \ref{sssec:iso}, and the two dimensional residual in Figure
\ref{fig:2Da}). The central surface brightness decline in {\bf E}
is likely to be caused by the bad column right at the center of the
galaxy.}
\label{fig:profs}
\end{figure*}

Isophotes were fitted to the images with the center of the ellipse,
the semiaxes $a$ and $b$, and the position angle as free parameters;
variations of ellipticity ($e$) and position angle ($pa$) with the
``circularised'' radius $\sqrt{ab}$ were allowed.  Isophotes were
derived using a version of the {\sc midas} command {\sc fit/ell3}
modified to better deal with steep gradients in the luminosity profile
in the innermost pixels (M{\o}ller, Stiavelli \& Zeilinger, 1995).
The profiles obtained were fitted with an exponential law, an \dv law,
and a linear combination of the two. In all cases, except for {\bf B},
we found that the objects were best fitted by the $r^{1/4}$ law. A
slightly better fit of the light profile of galaxy {\bf B} was obtained
by adding a small exponential component to the \dv law, suggesting that the
bump on the residual might be due to a fainter disk component (see
Figures \ref{fig:profs} and \ref{fig:2Da}).  The fit was done using
the least squares fitting software used by Carollo et al.\
\shortcite{MCetal}.  For each galaxy we used a specific PSF calculated at its
position.  In Figure~\ref{fig:profs} we show the
best-fit \dv laws convolved with the PSFs, superimposed on the data
points.

\subsubsection{Two-dimensional fits}

This technique consists in fitting a two-dimensional model of an
elliptical galaxy, as seen on the CCD chip, to the WFPC2 image. The
model galaxy has an \dv luminosity profile, with fixed position angle
and ellipticity (we used the value derived from the isophotal fit near
the effective radius). Each fit has four free parameters: the
effective surface brightness, the effective radius, and the position
of the centre.  We have two sets of results, depending on how the fit
was performed (see (iv) below). The software we developed works as
follows:

\begin{enumerate}

\item The code generates a 2D projected image of the galaxy using the
\dv law on a subsampled grid.  The pixel size is chosen to be
approximately ten times smaller than the galaxy effective radius in
order to make discretisation problems negligible.

\item The resulting ``ideal'' model is rebinned to the WFPC2 chip scale. 

\item The rebinned model is convolved with the Tiny Tim \cite{TT4} PSF. 
The Tiny Tim (non resampled) PSF includes the effects of diffraction
(i.e. of the telescope and camera optics) and the Pixel
Response Function (i.e. the spread caused by electron diffusion in the
WFPC2 CCDs). 

\item This model is fitted to the data either by least $\chi^2$ or by
least squares, using a simplex algorithm to find the minimum ({\sc
amoeba}, Press et al.~1992). 

\end{enumerate}

\begin{figure*}
\begin{center}

A

\mbox{
\mbox{\epsfysize=3cm \epsfbox{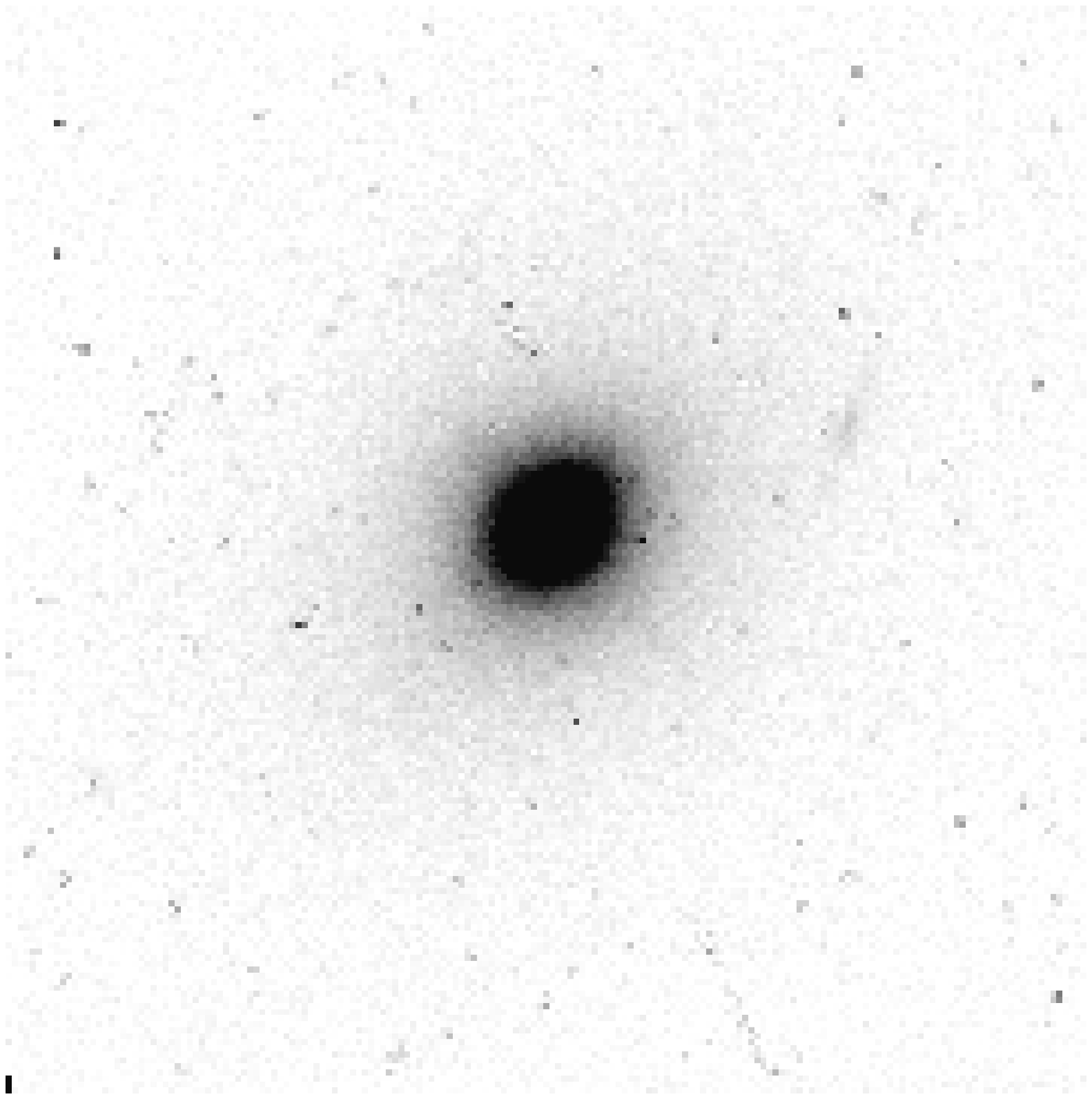}}
\mbox{\epsfysize=3cm \epsfbox{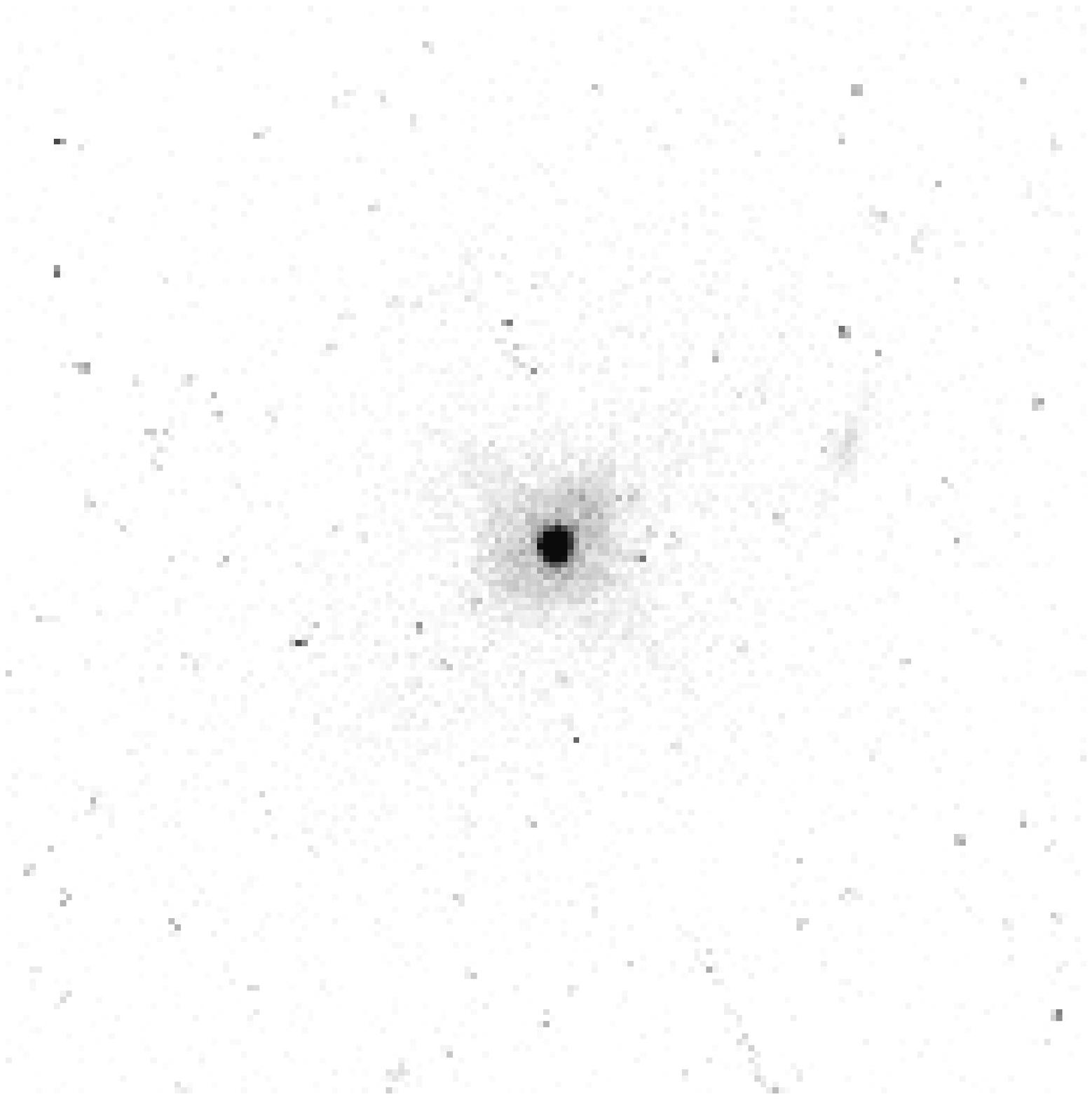}}
\mbox{\epsfysize=3cm \epsfbox{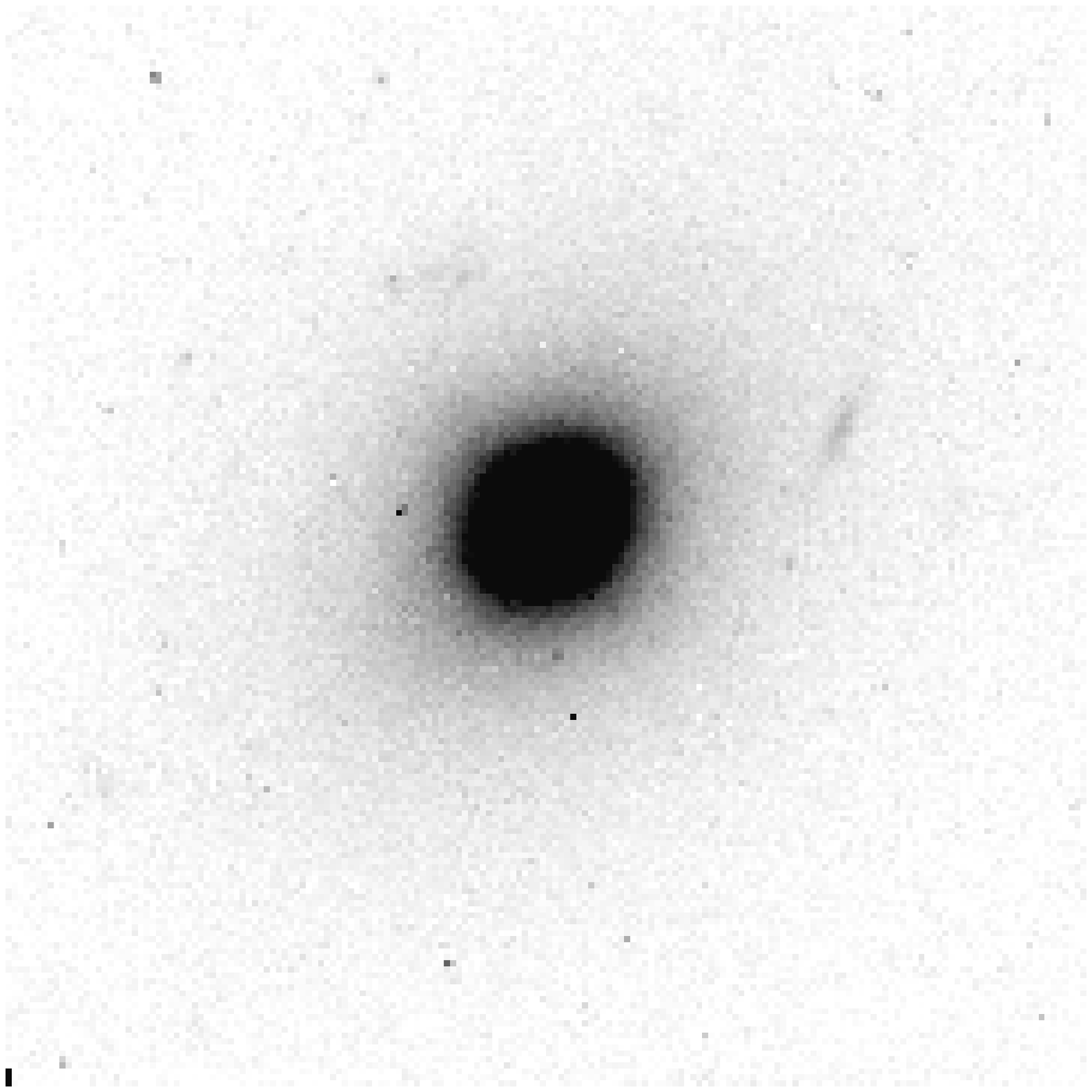}}
\mbox{\epsfysize=3cm \epsfbox{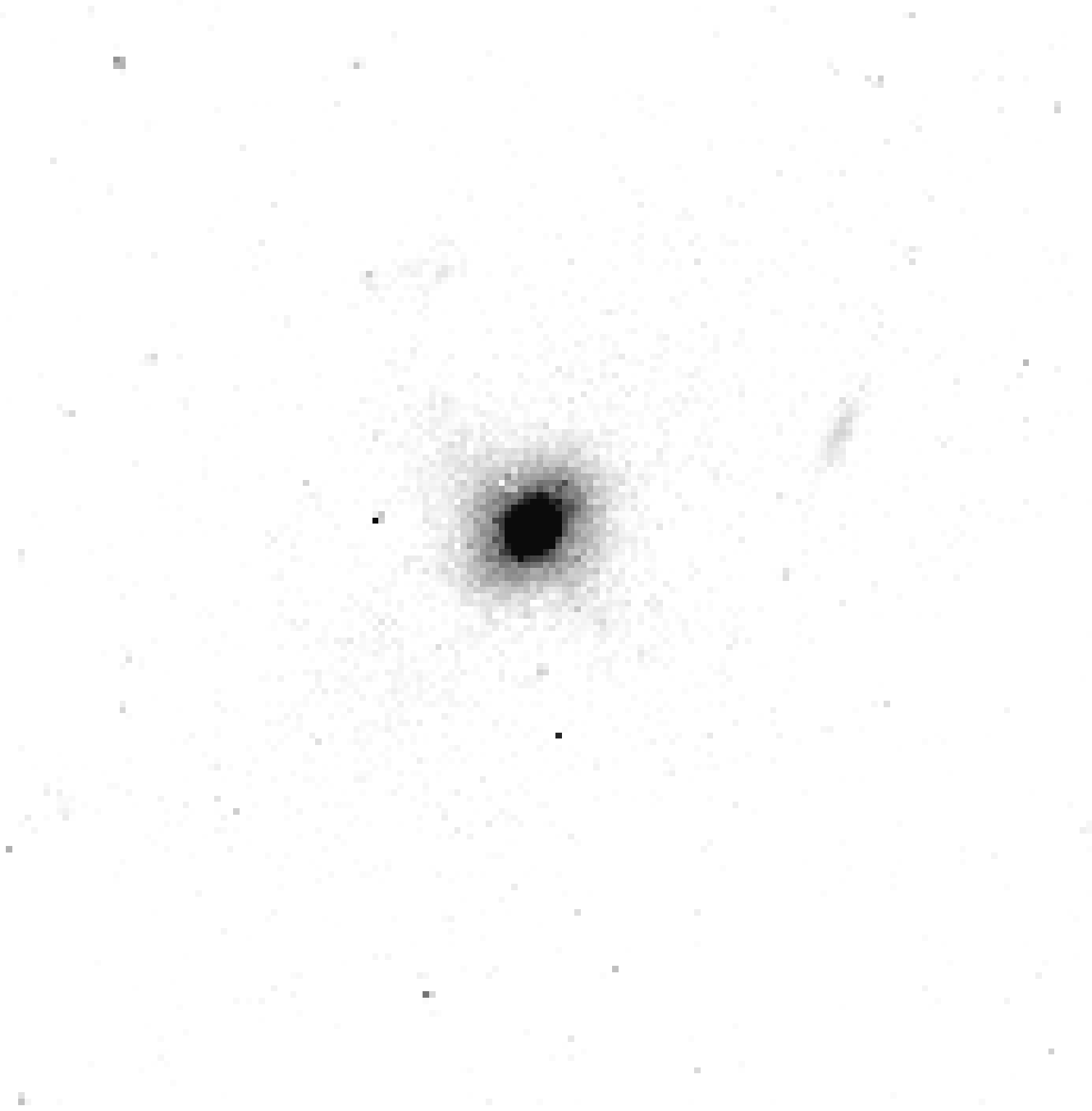}}
}

D

\mbox{
\mbox{\epsfysize=3cm \epsfbox{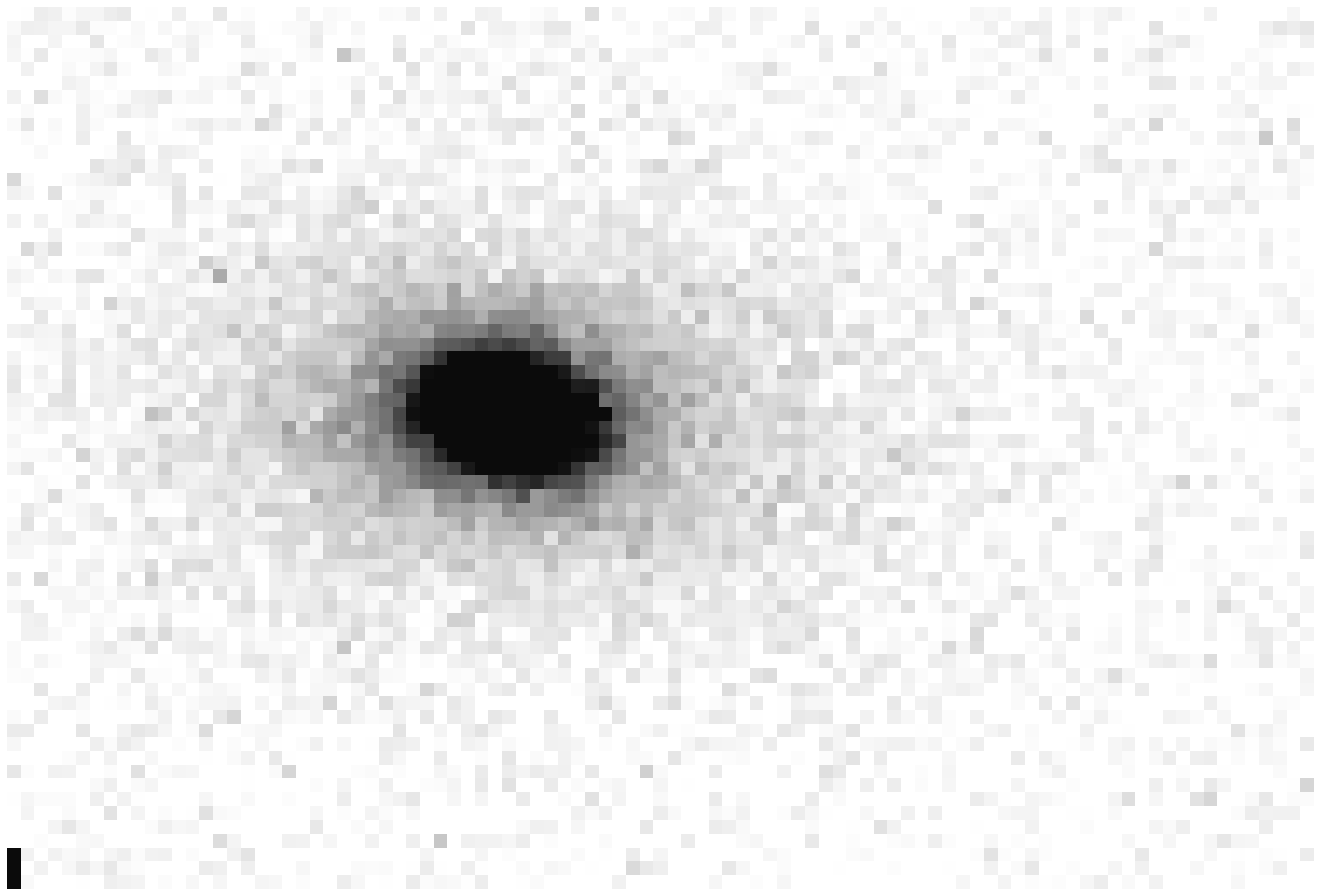}}
\mbox{\epsfysize=3cm \epsfbox{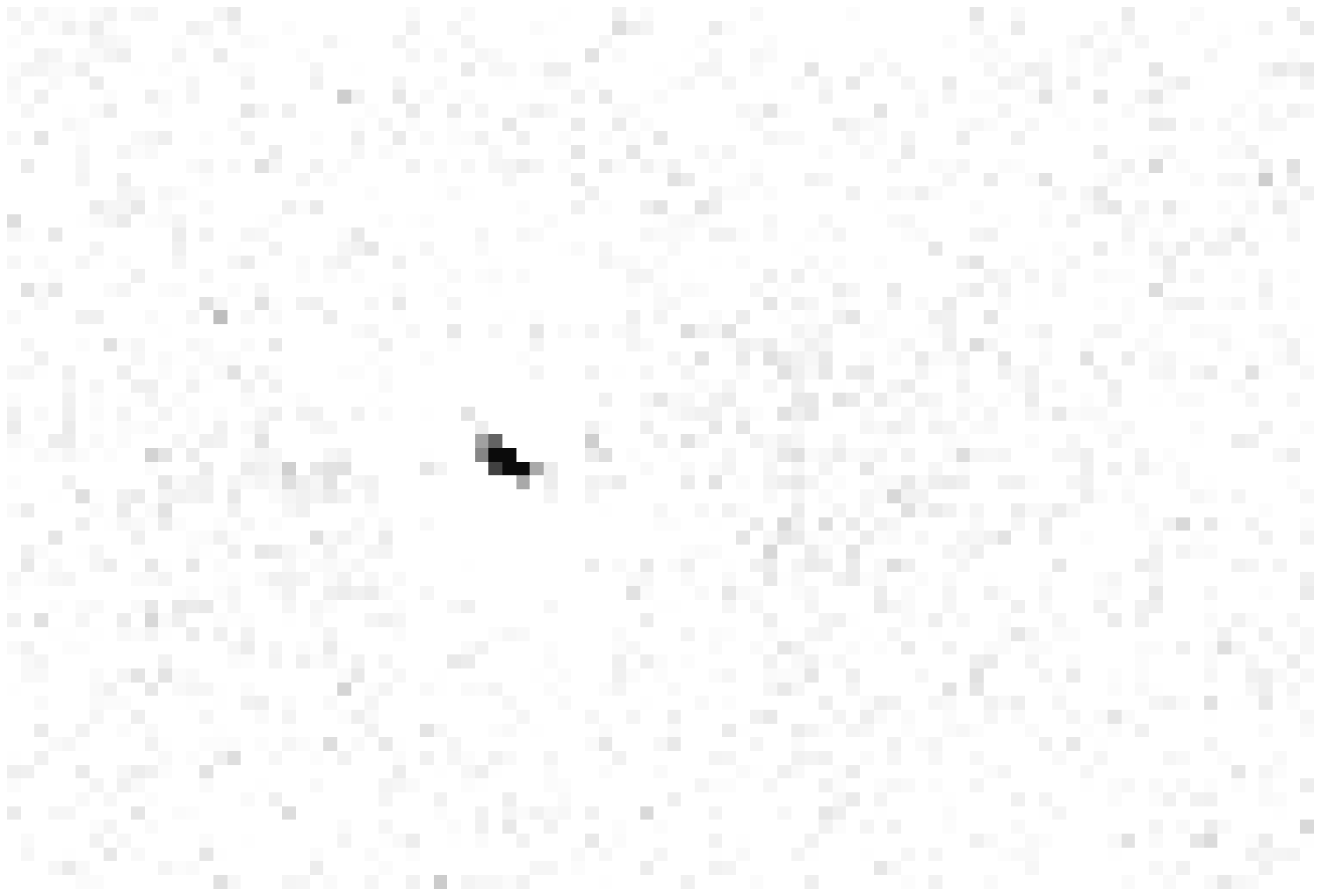}}
\mbox{\epsfysize=3cm \epsfbox{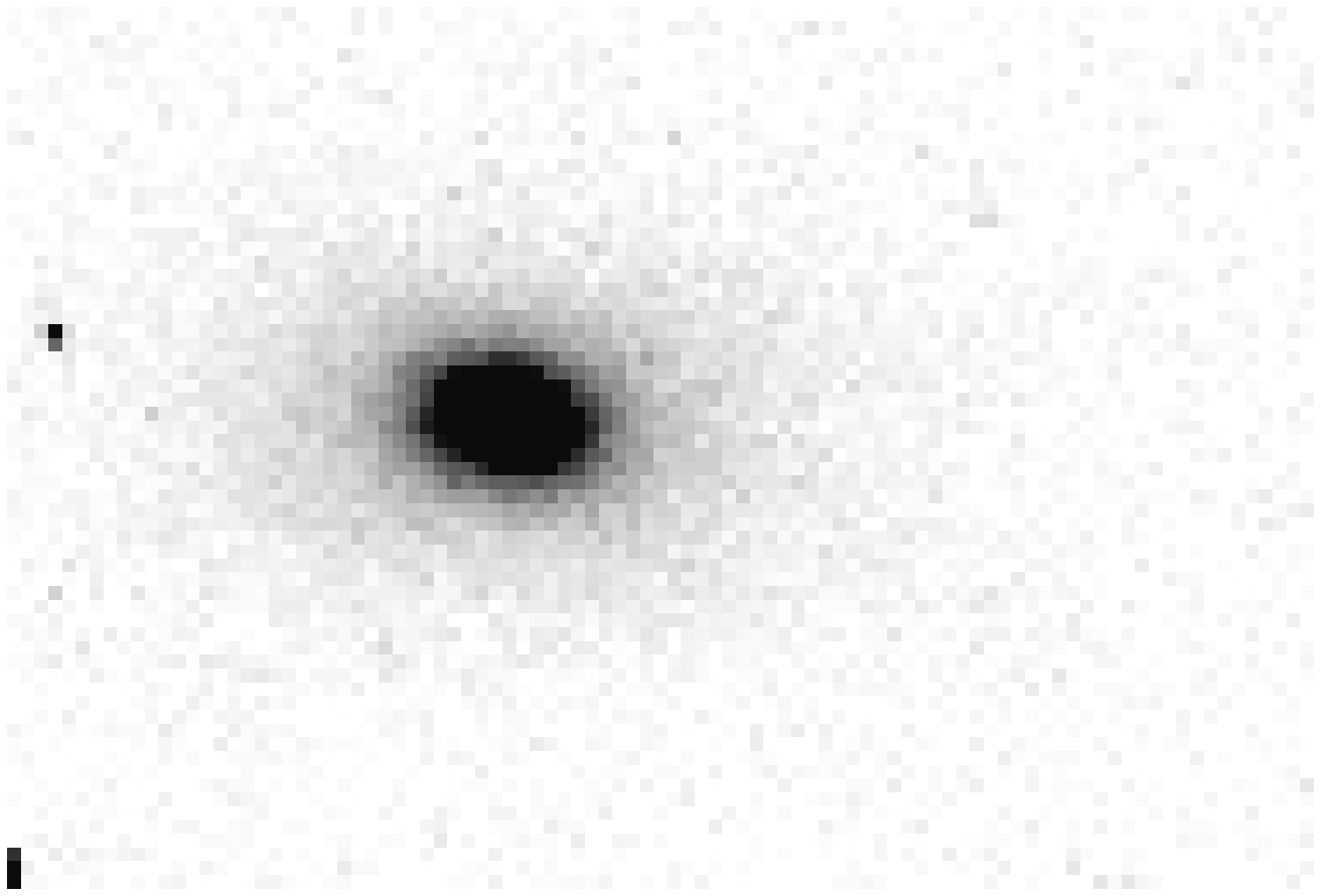}}
\mbox{\epsfysize=3cm \epsfbox{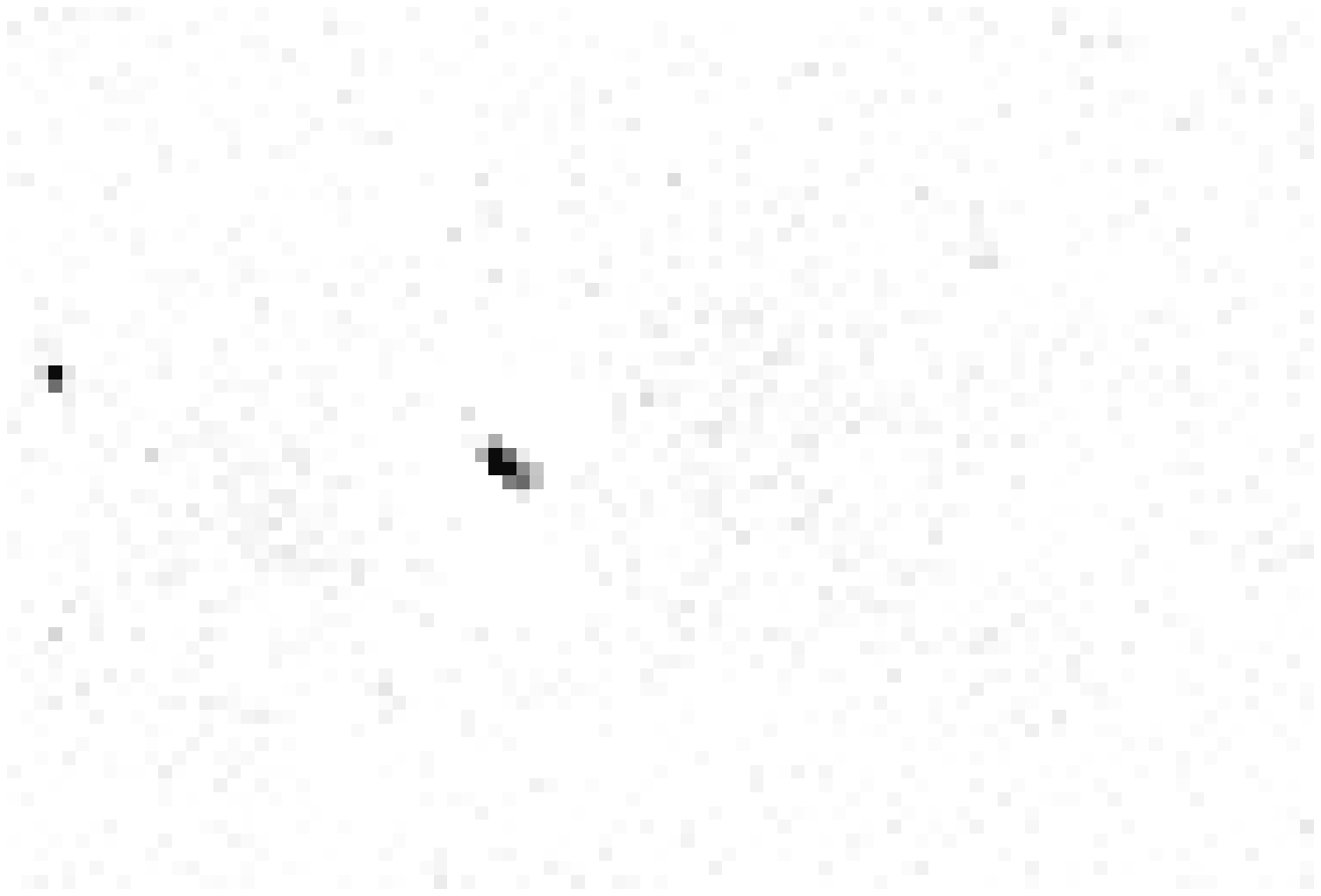}}}

E

\mbox{
\mbox{\epsfysize=3cm \epsfbox{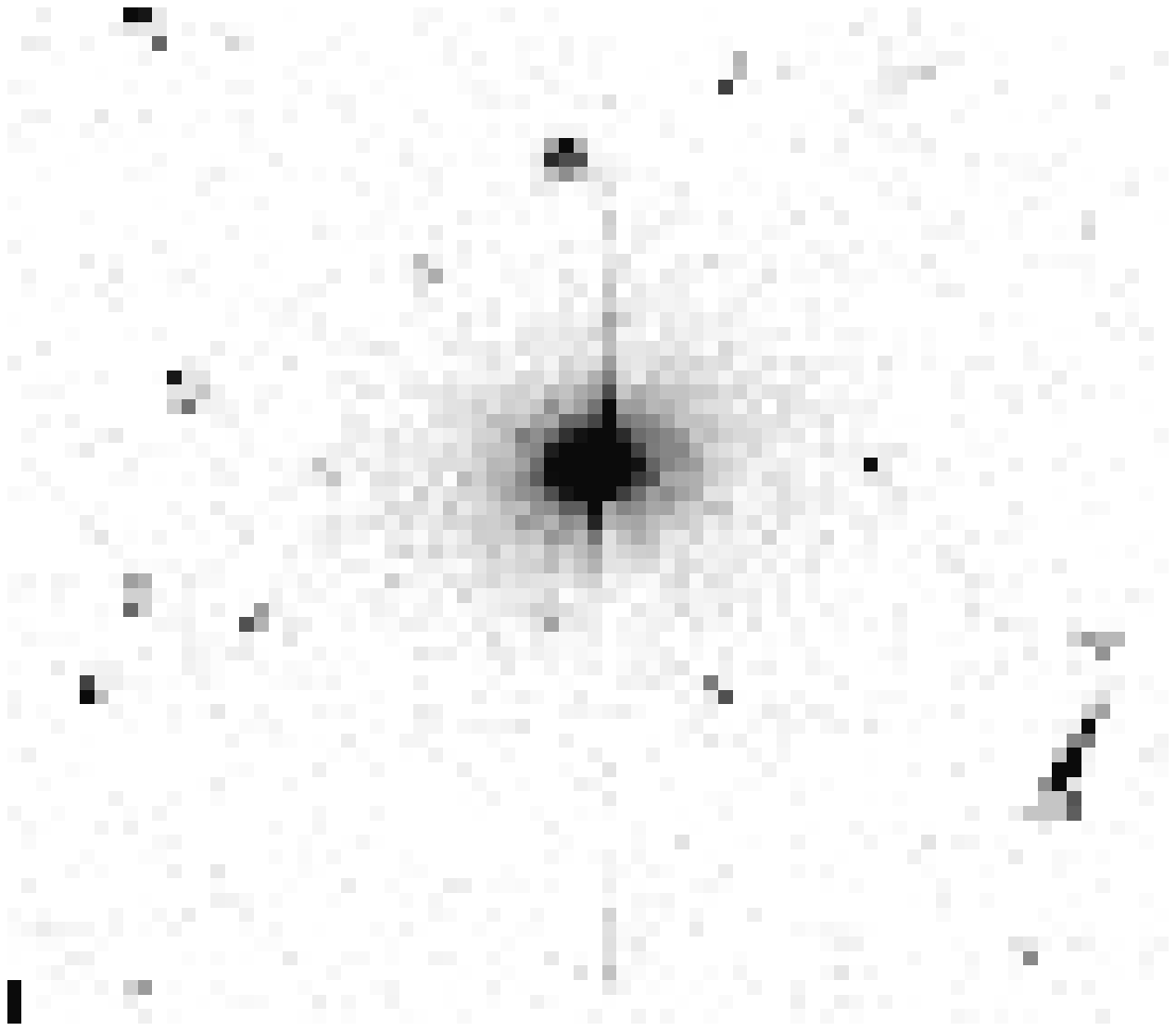}}
\mbox{\epsfysize=3cm \epsfbox{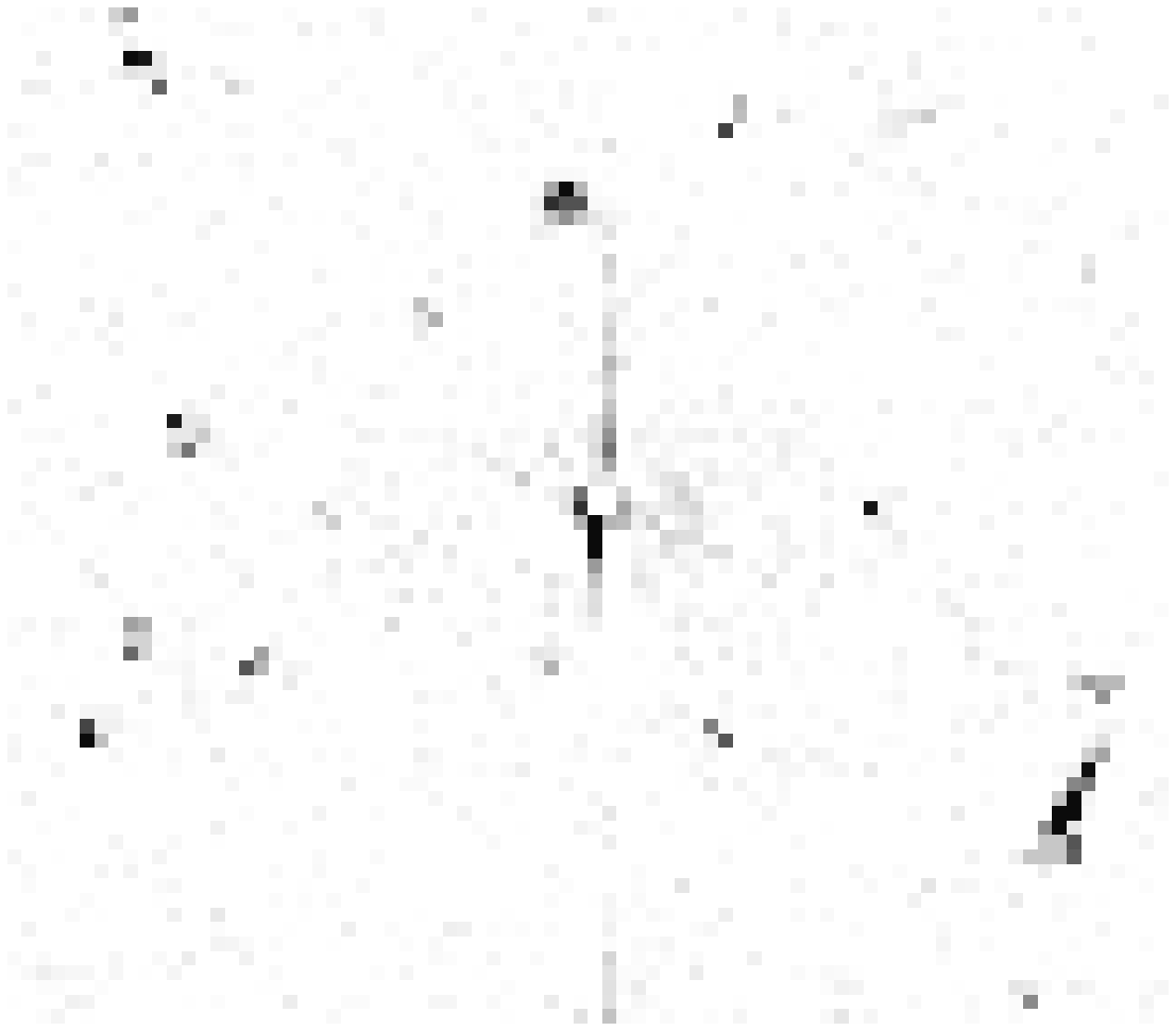}}
\mbox{\epsfysize=3cm \epsfbox{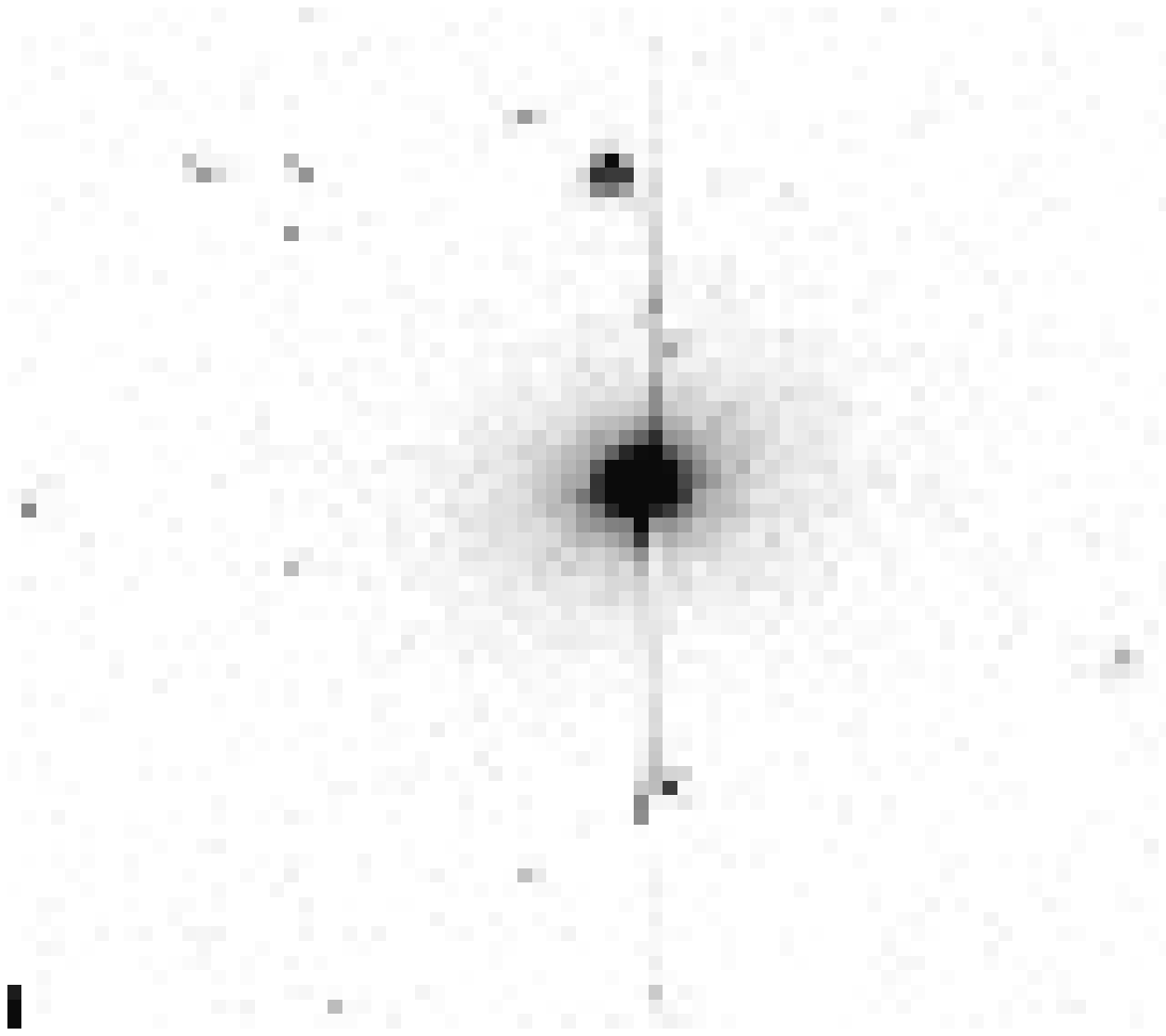}}
\mbox{\epsfysize=3cm \epsfbox{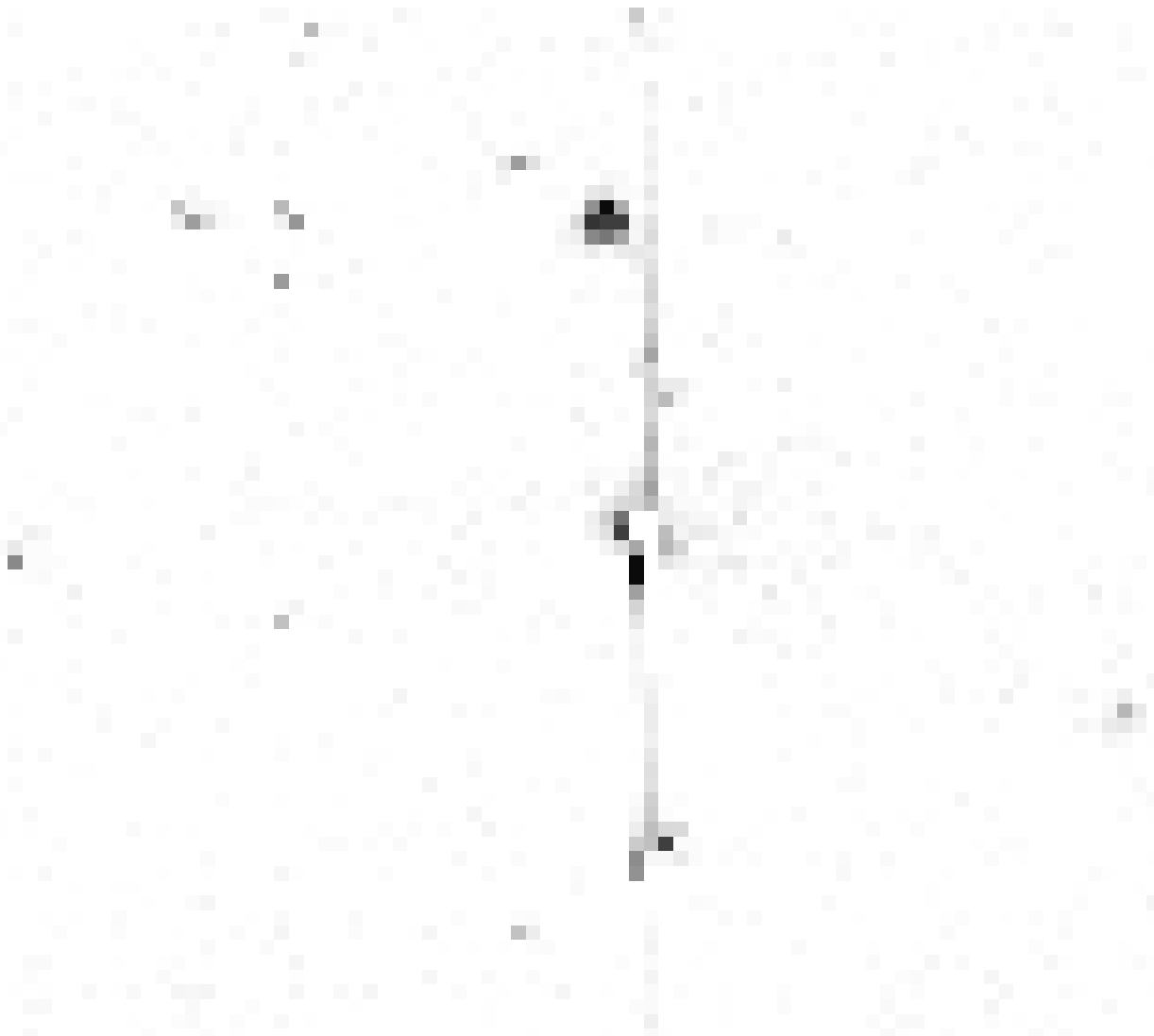}}}

F

\mbox{
\mbox{\epsfysize=3cm \epsfbox{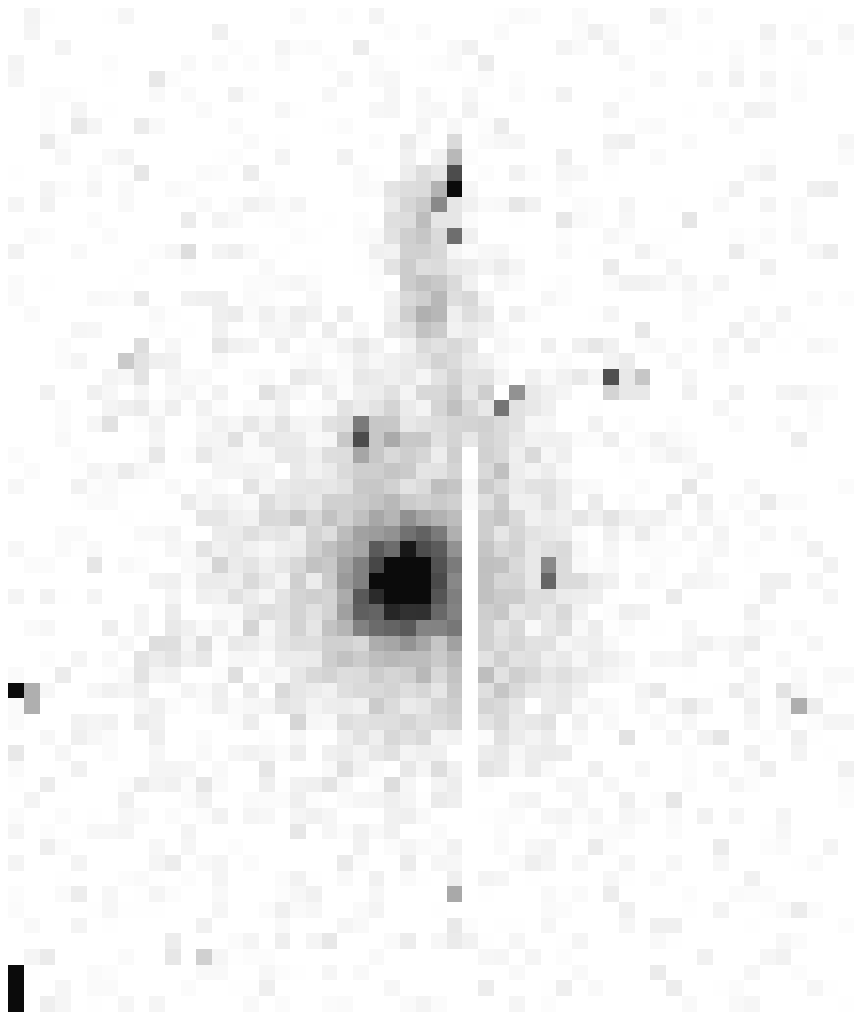}}
\mbox{\epsfysize=3cm \epsfbox{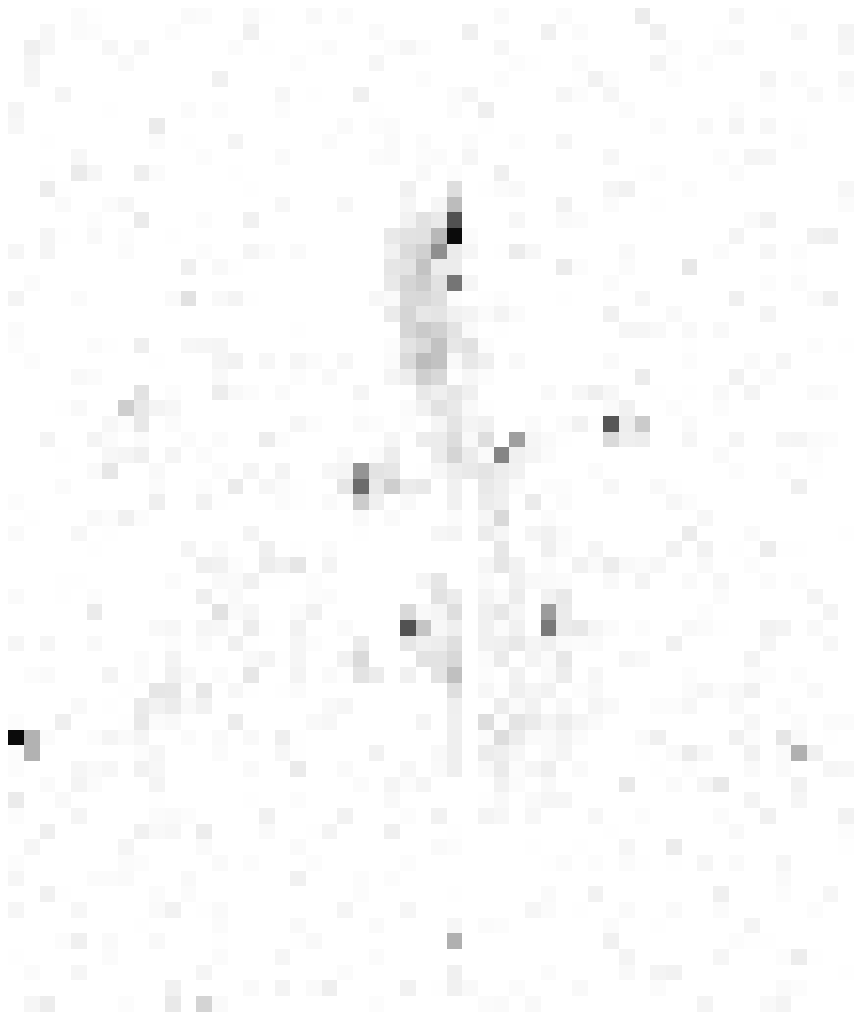}}
\mbox{\epsfysize=3cm \epsfbox{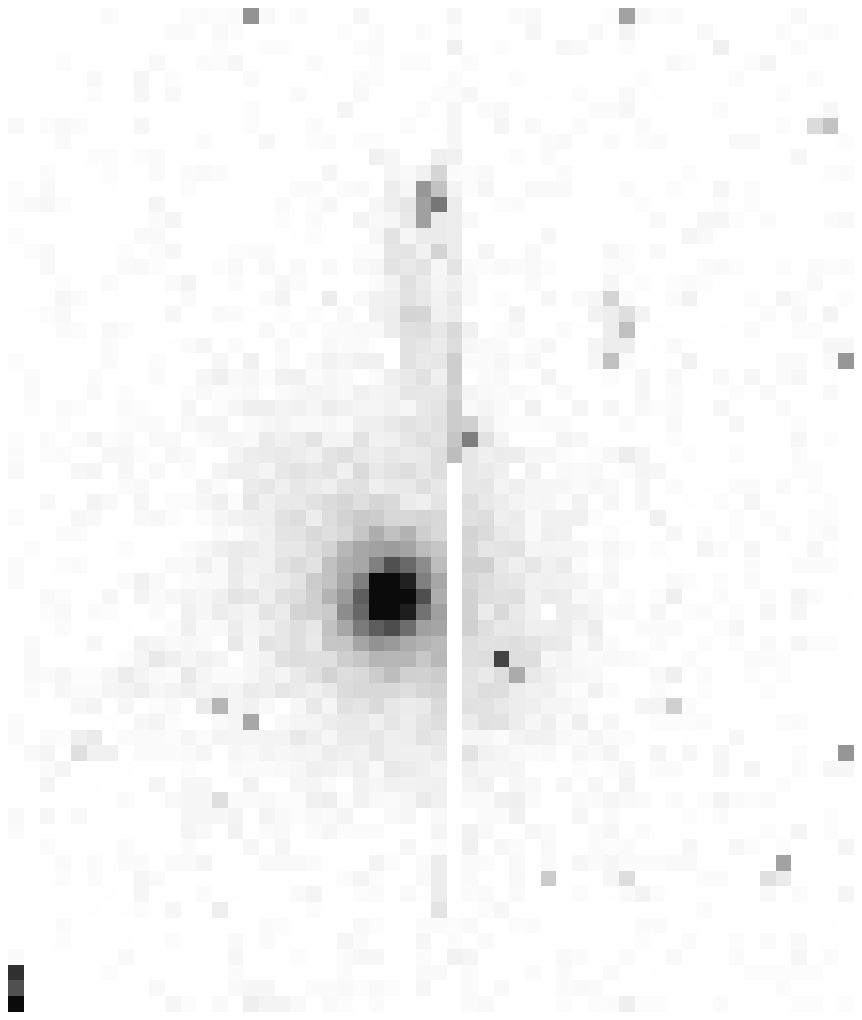}}
\mbox{\epsfysize=3cm \epsfbox{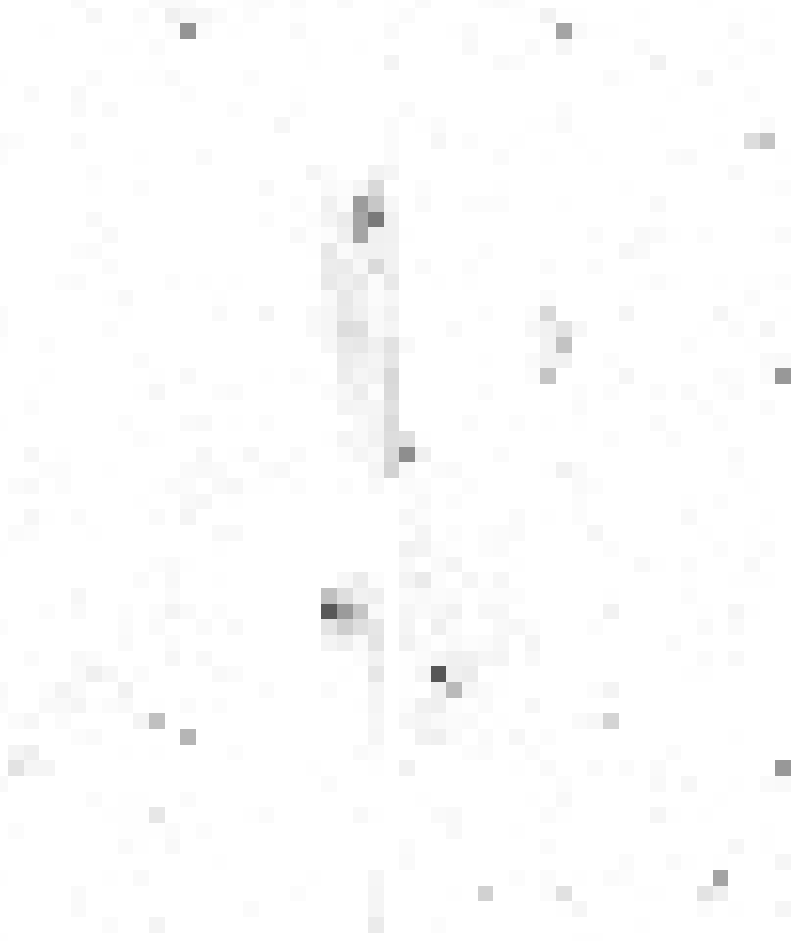}}}

GH

\mbox{
\mbox{\epsfysize=3cm \epsfbox{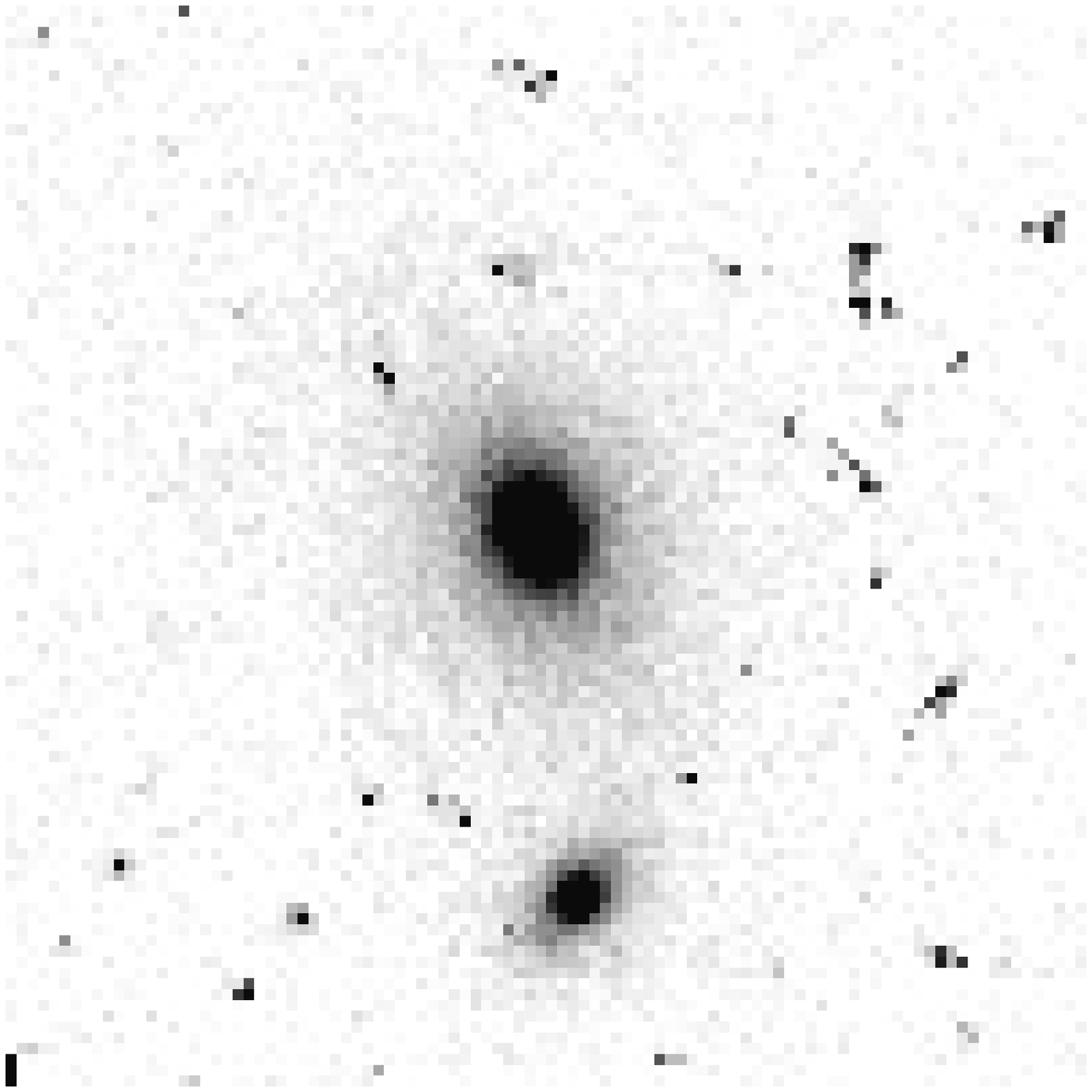}}
\mbox{\epsfysize=3cm \epsfbox{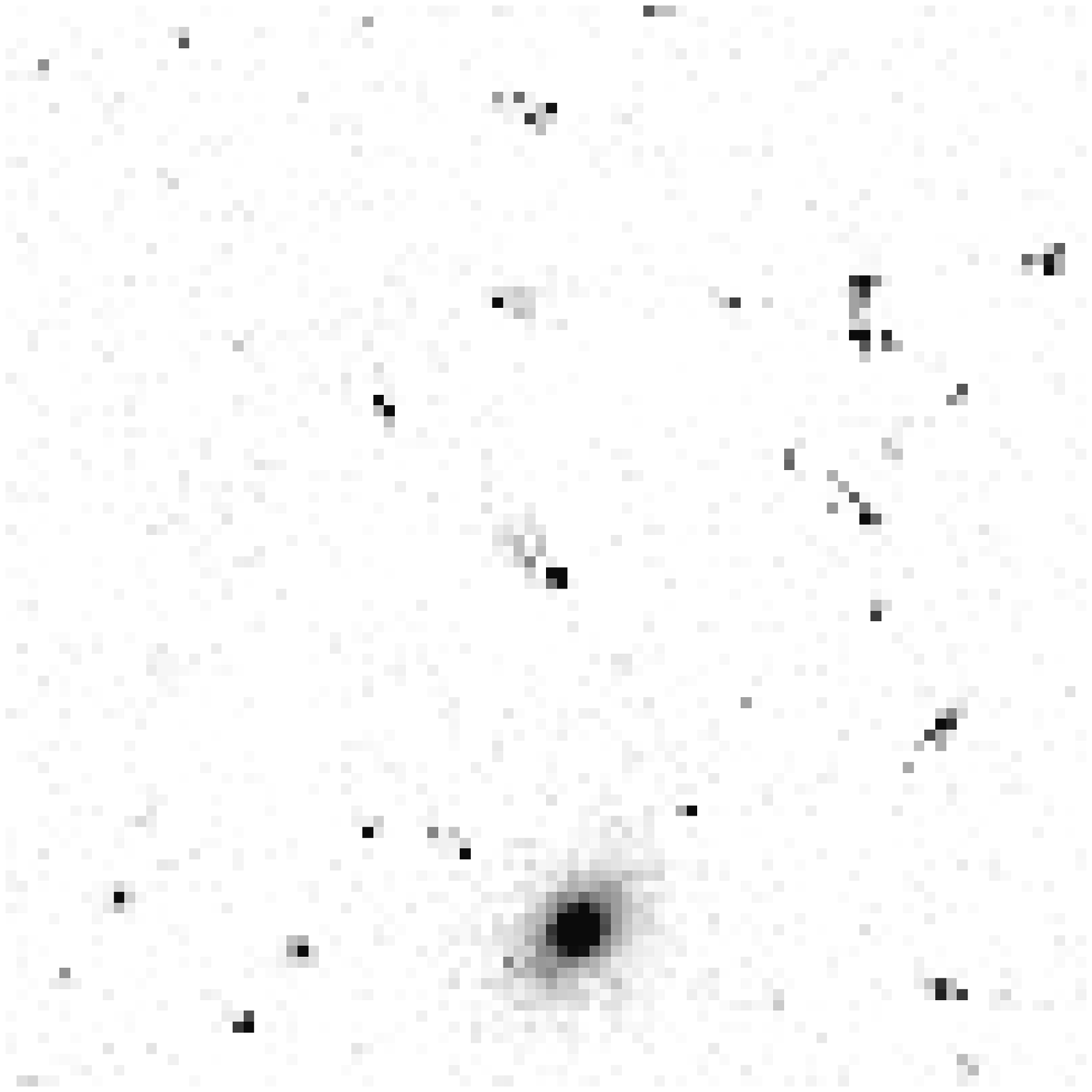}}
\mbox{\epsfysize=3cm \epsfbox{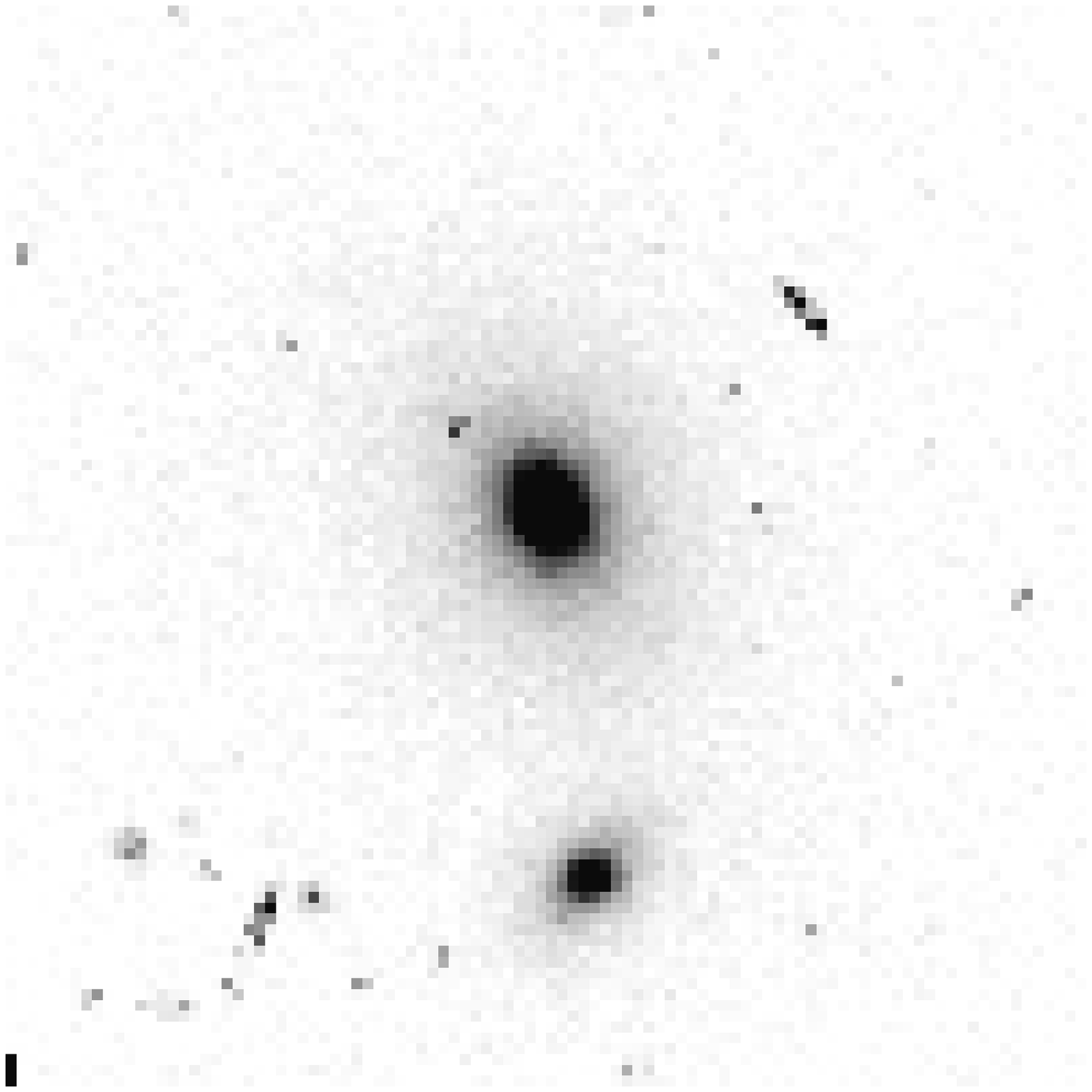}}
\mbox{\epsfysize=3cm \epsfbox{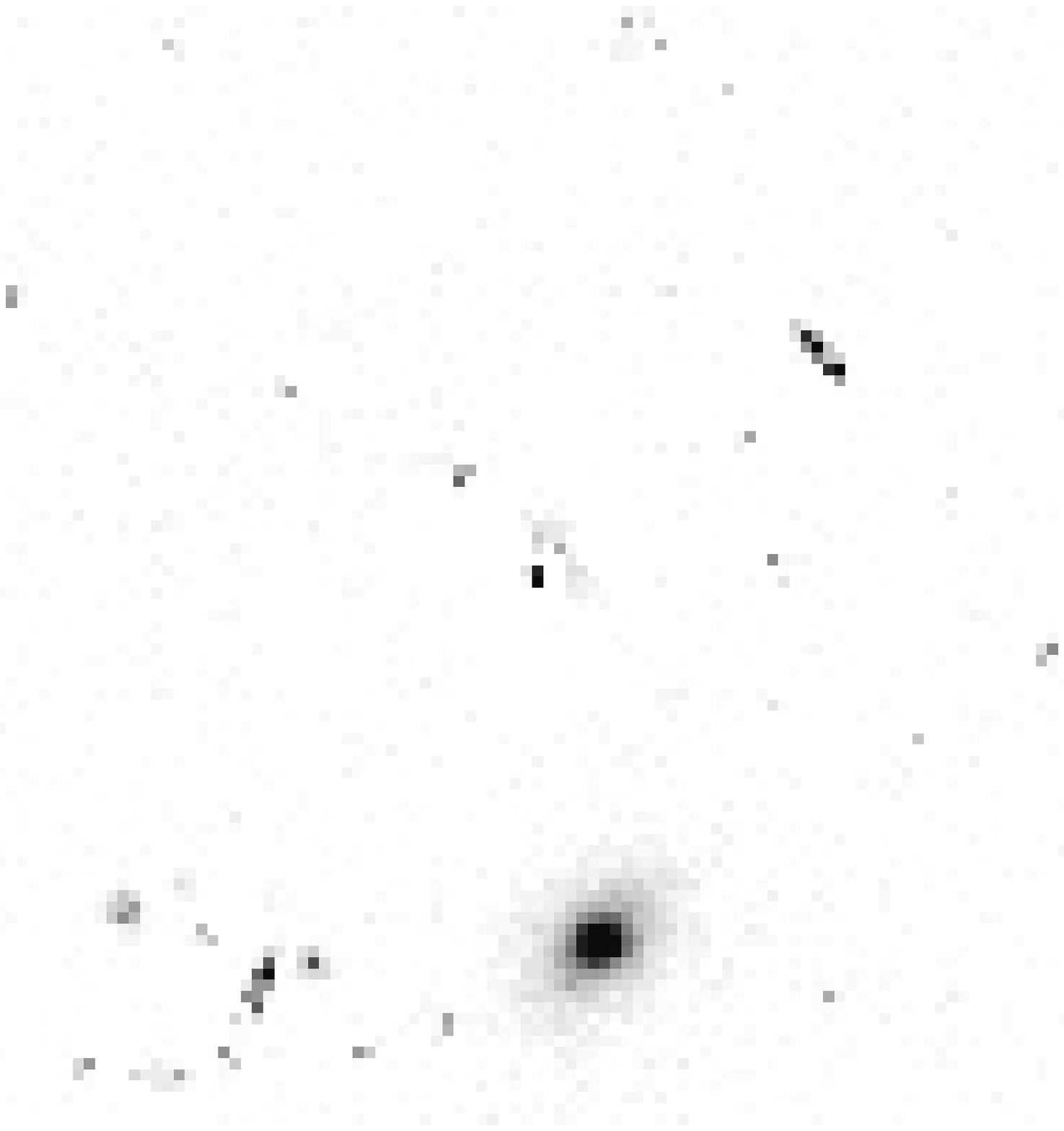}}}

I

\mbox{
\mbox{\epsfysize=3cm \epsfbox{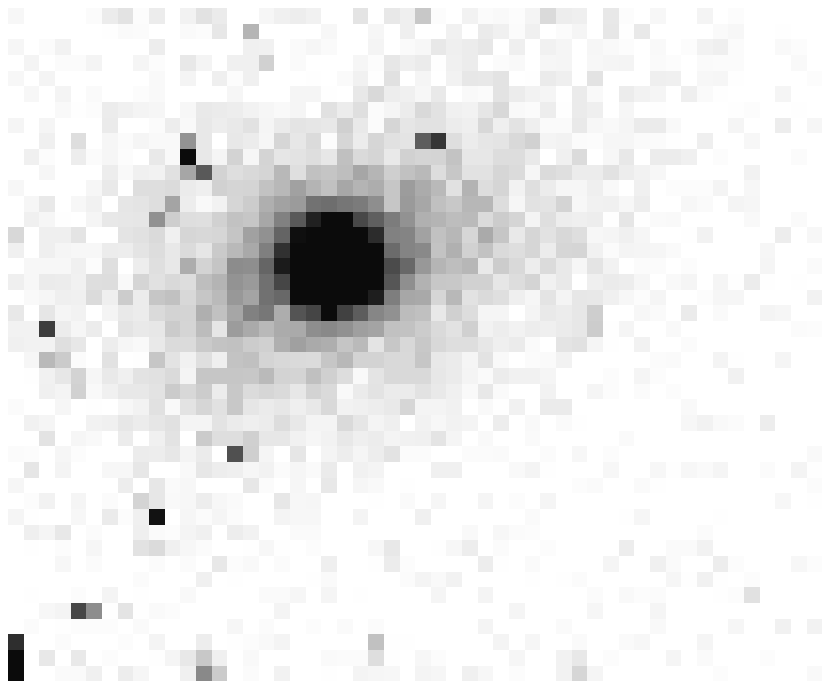}}
\mbox{\epsfysize=3cm \epsfbox{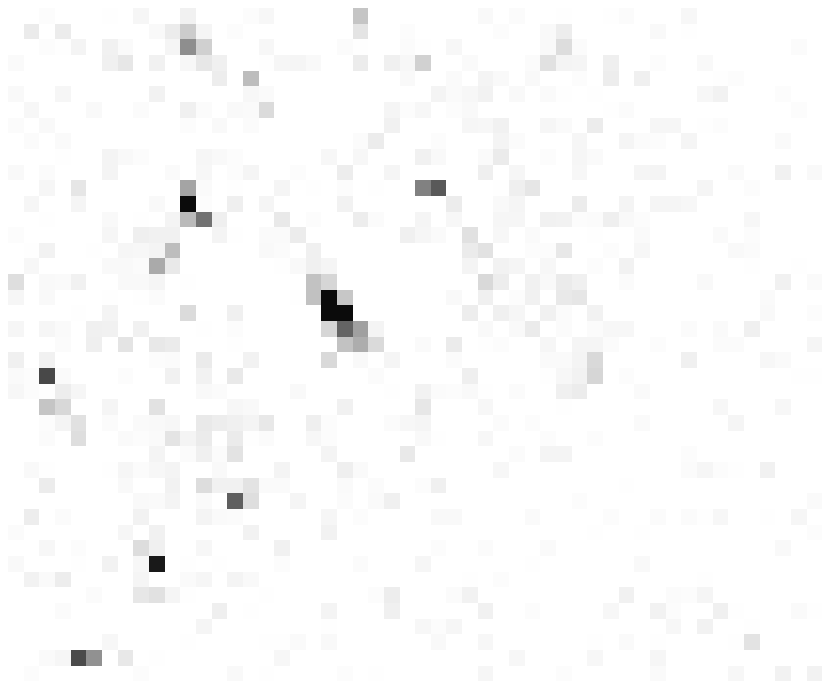}}
\mbox{\epsfysize=3cm \epsfbox{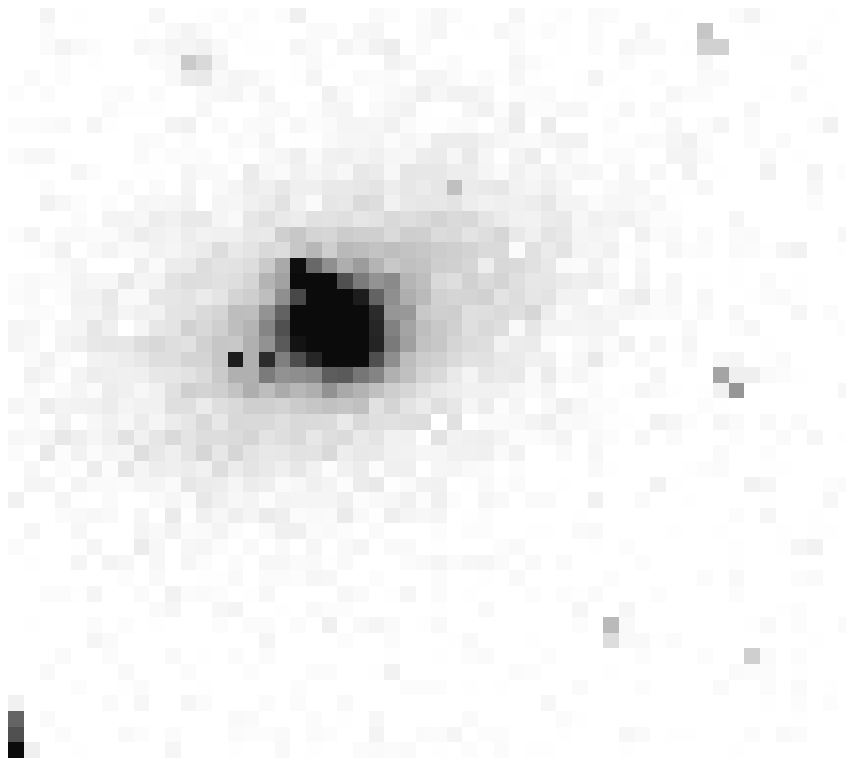}}
\mbox{\epsfysize=3cm \epsfbox{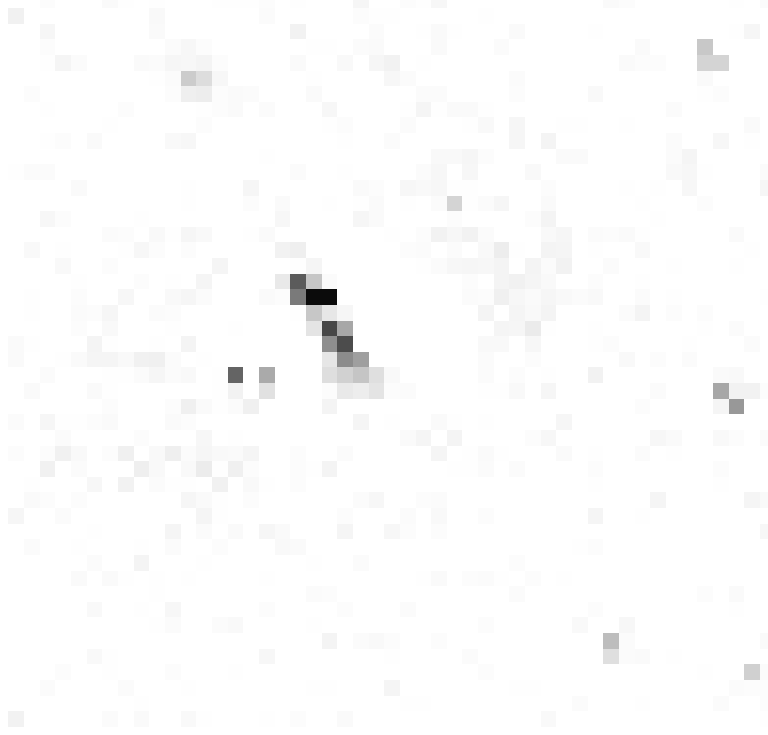}}}

\end{center}
\caption{WFPC2 images and 2D fit residuals for the galaxies with
measured kinematics: {\bf A},{\bf D}, {\bf E}, {\bf F}, {\bf G}, {\bf
I}. From the left to the right we display the F606W image, the F606W
residual, the F814W image and the F814W residual.  The images (and
their sizes) are the ones used for the 2D fits, see
Table~1.  Galaxies {\bf A}, {\bf D} show a residual in the
innermost pixels which is in agreement with the deviation from the \dv
law in the luminosity profile.  The bad column in the centre of {\bf
E} is clearly evident in the residual. {\bf F} shows a peculiar pot
handle structure just to the upper left of the bad column.}
\label{fig:2D}
\end{figure*}

\begin{figure*}
\begin{center}

B

\mbox{
\mbox{\epsfysize=3cm \epsfbox{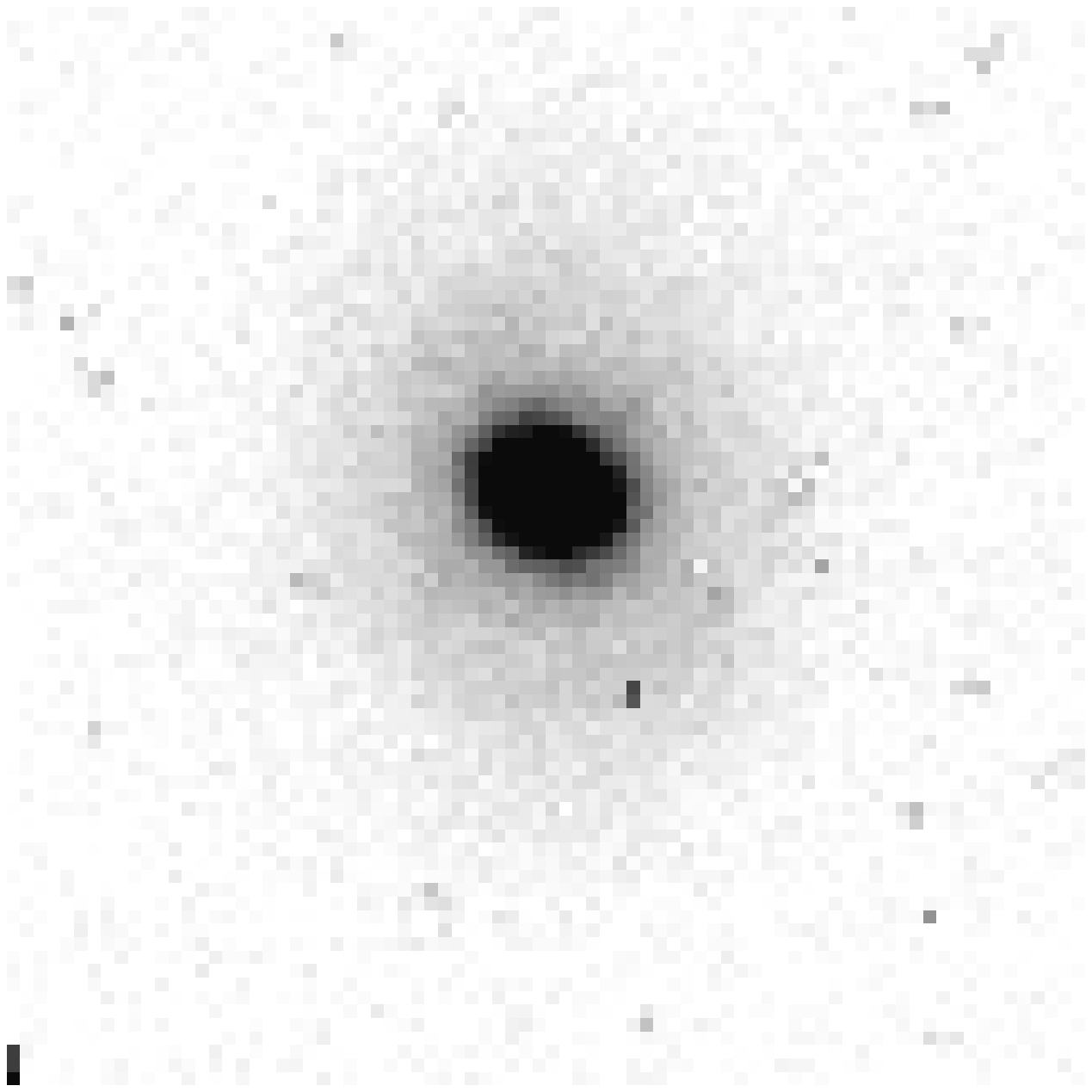}}
\mbox{\epsfysize=3cm \epsfbox{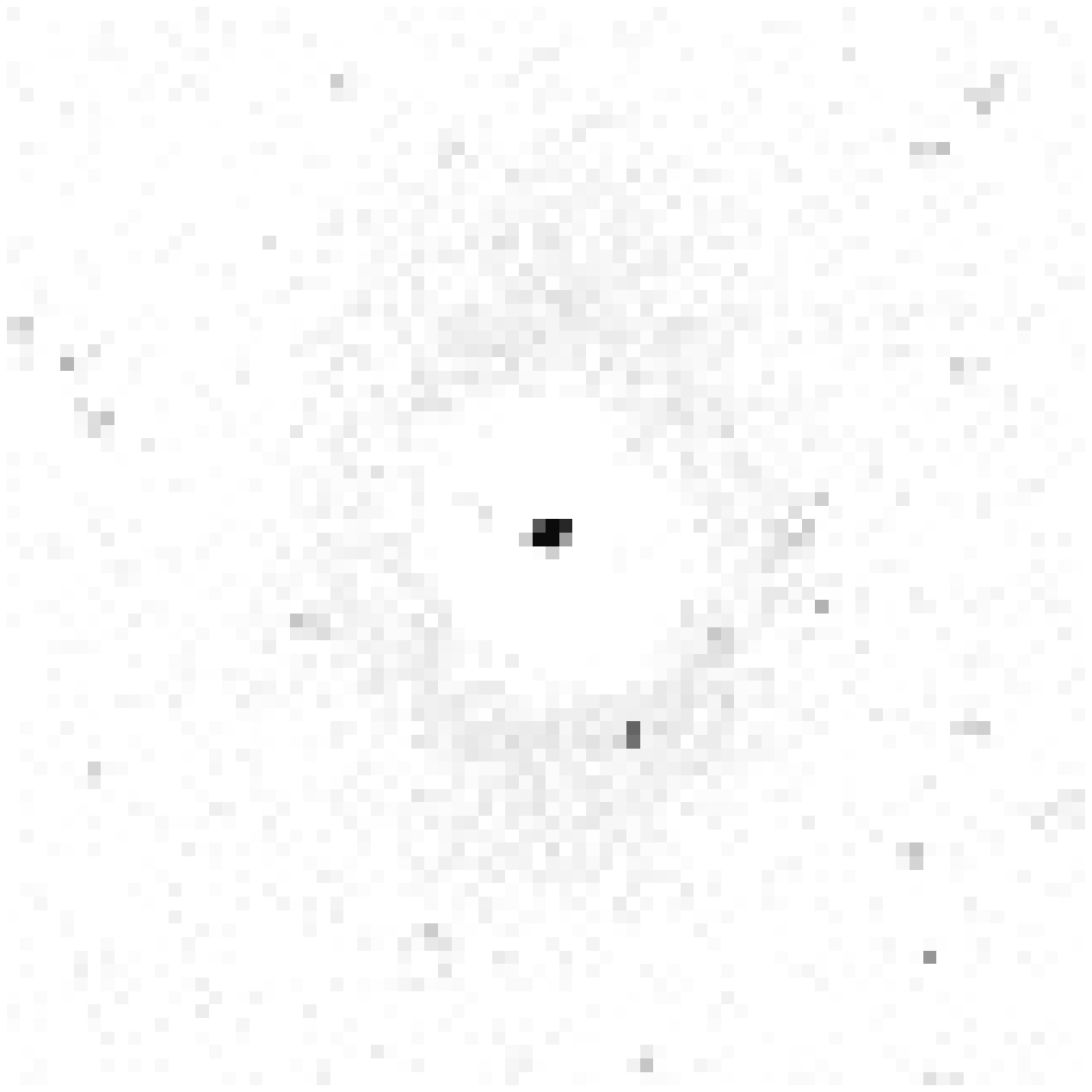}}
\mbox{\epsfysize=3cm \epsfbox{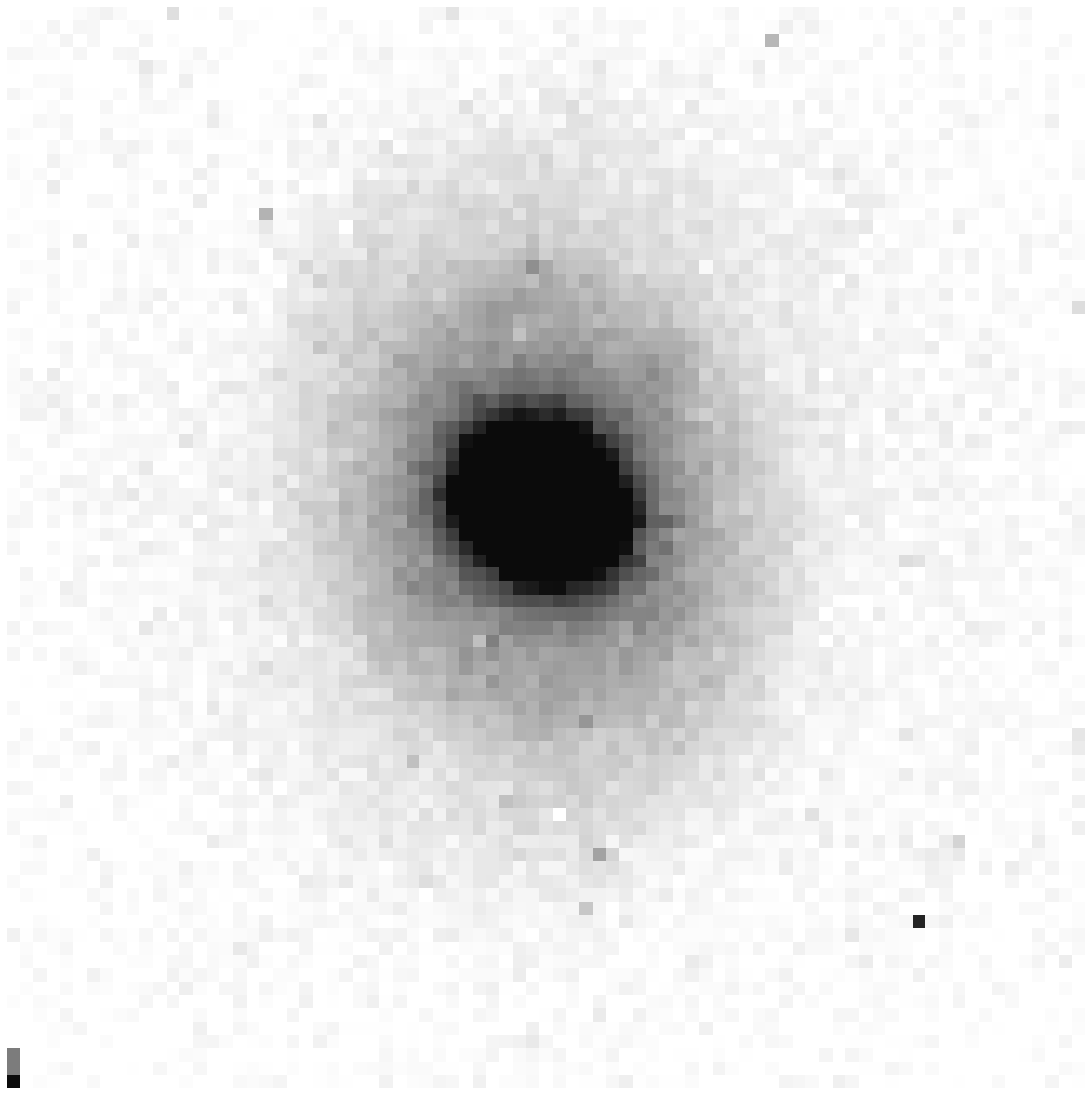}}
\mbox{\epsfysize=3cm \epsfbox{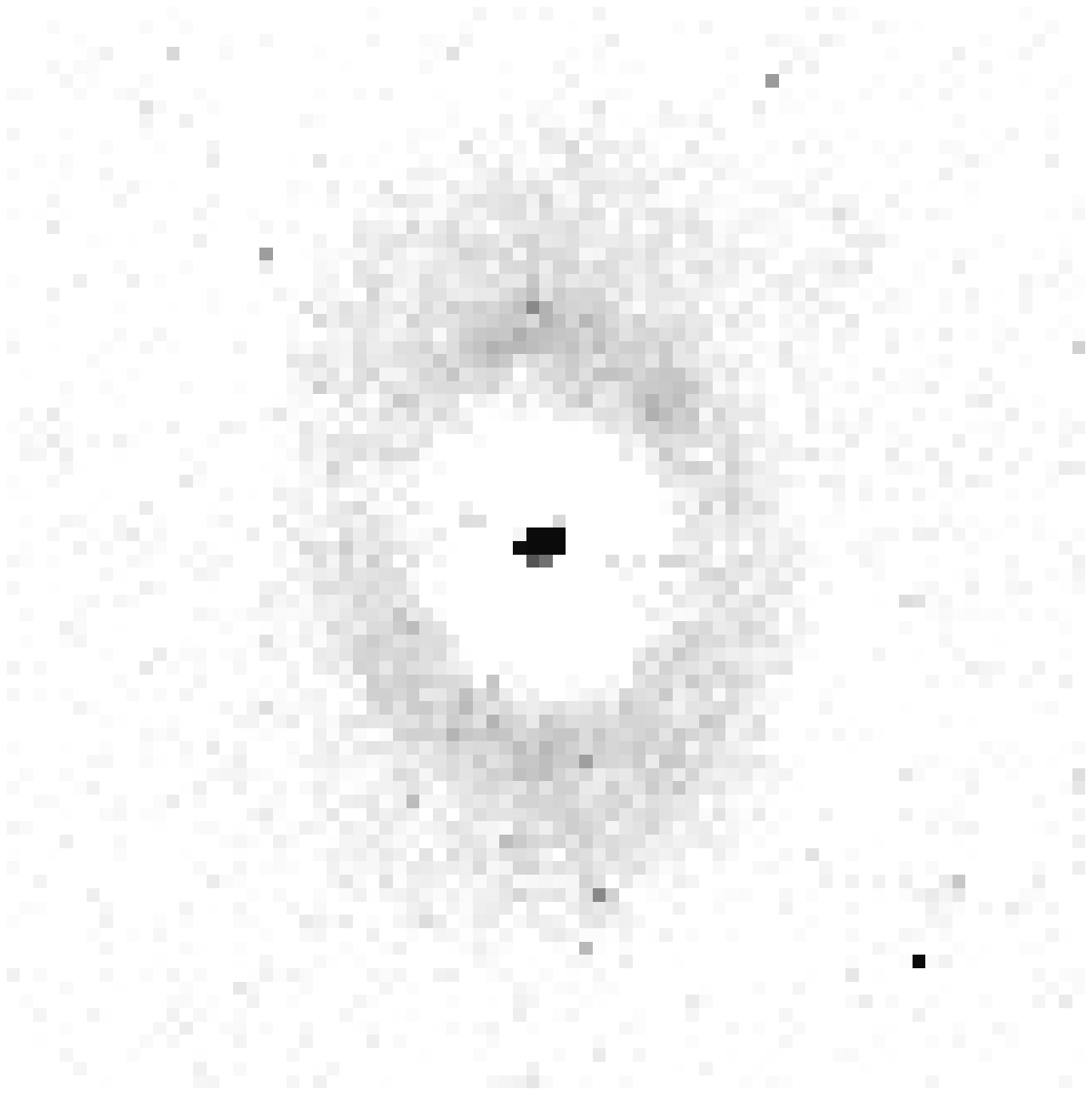}}}

O

\mbox{
\mbox{\epsfysize=3cm \epsfbox{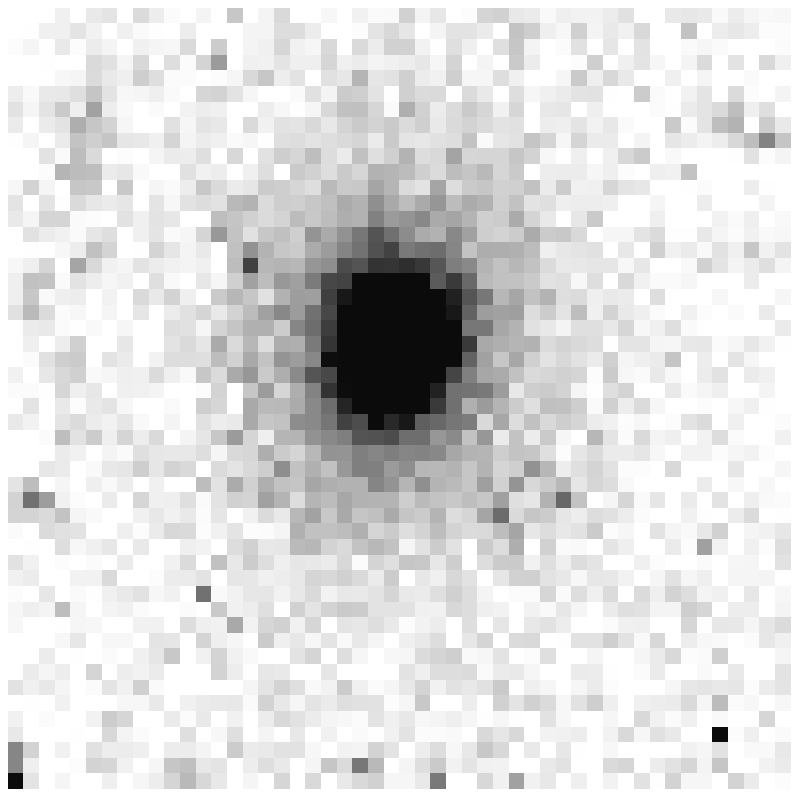}}
\mbox{\epsfysize=3cm \epsfbox{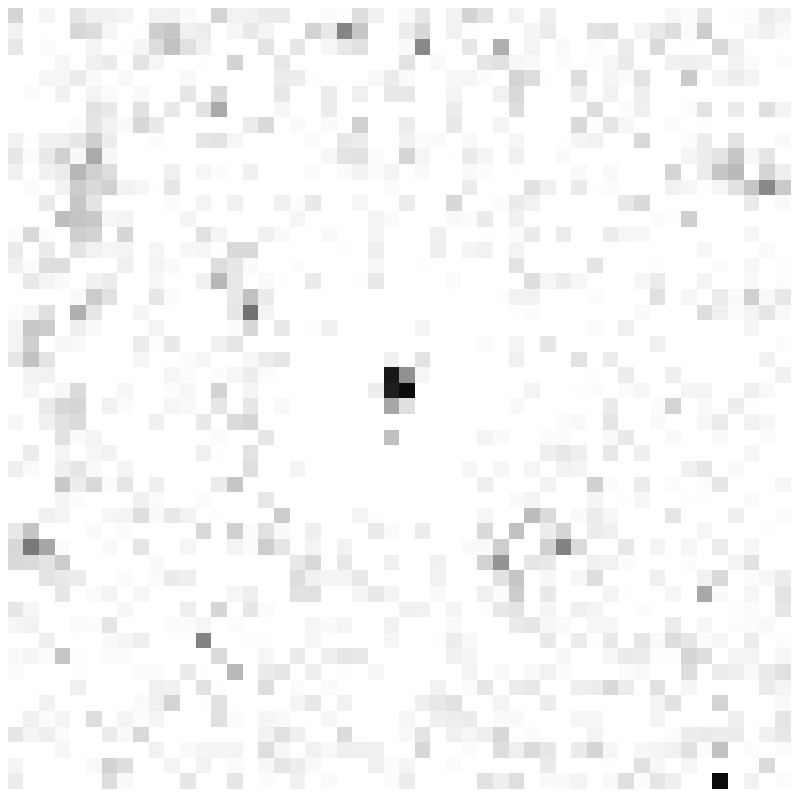}}
\mbox{\epsfysize=3cm \epsfbox{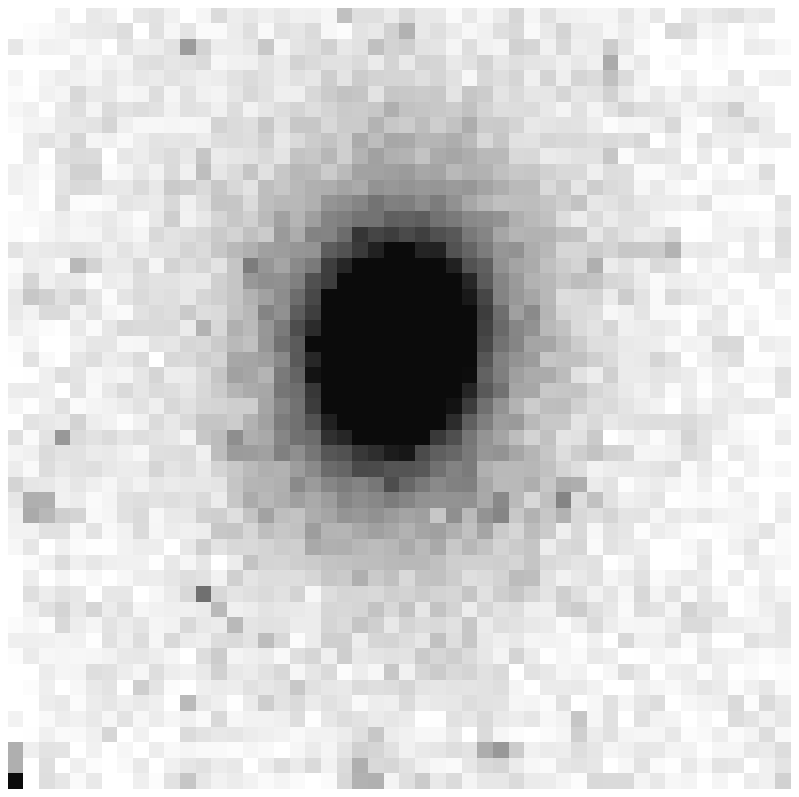}}
\mbox{\epsfysize=3cm \epsfbox{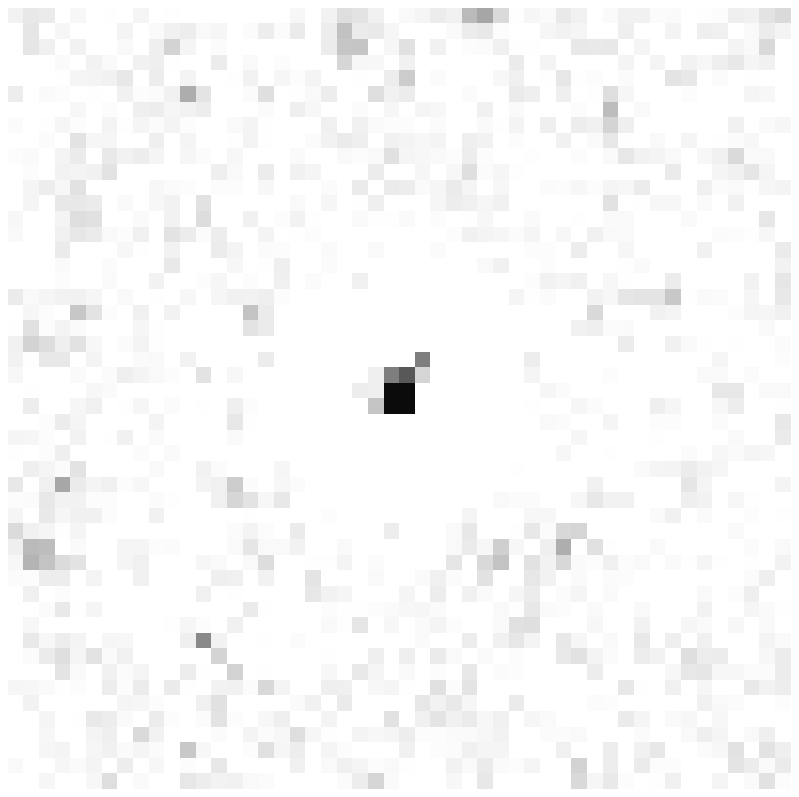}}}

P

\mbox{
\mbox{\epsfysize=3cm \epsfbox{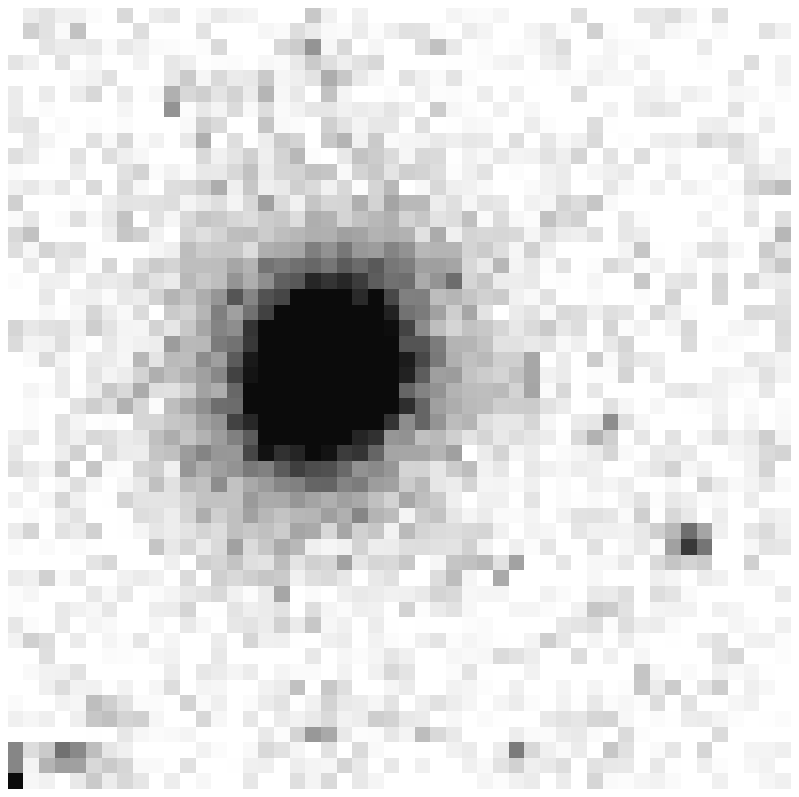}}
\mbox{\epsfysize=3cm \epsfbox{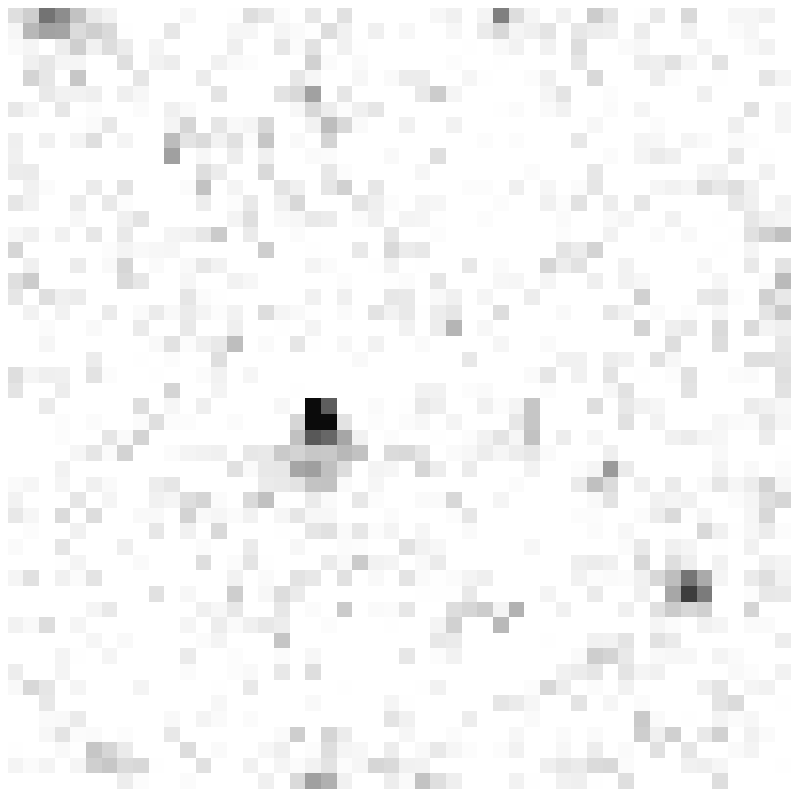}}
\mbox{\epsfysize=3cm \epsfbox{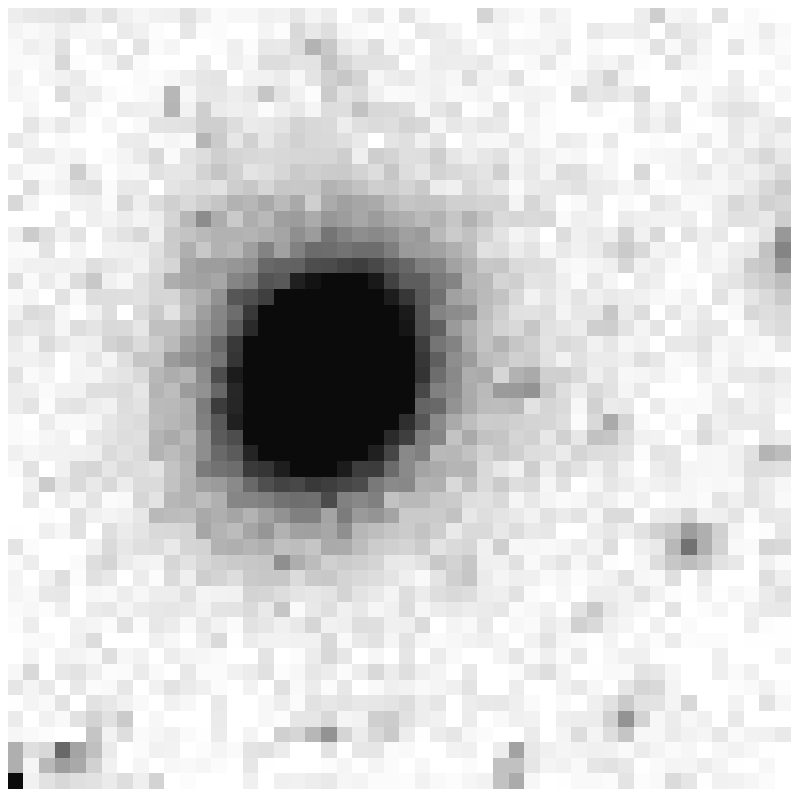}}
\mbox{\epsfysize=3cm \epsfbox{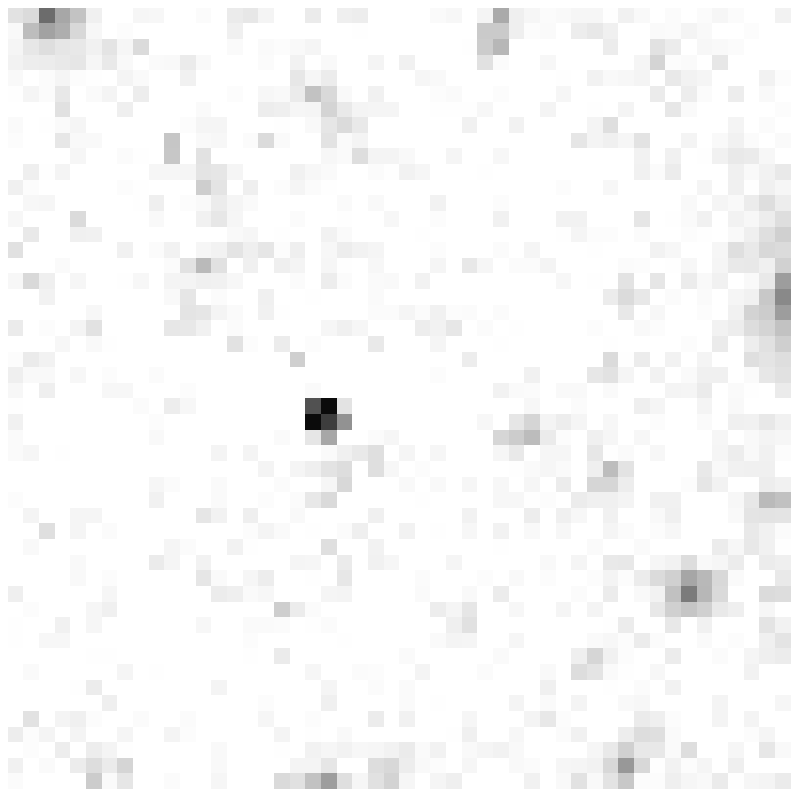}}}

Q

\mbox{
\mbox{\epsfysize=3cm \epsfbox{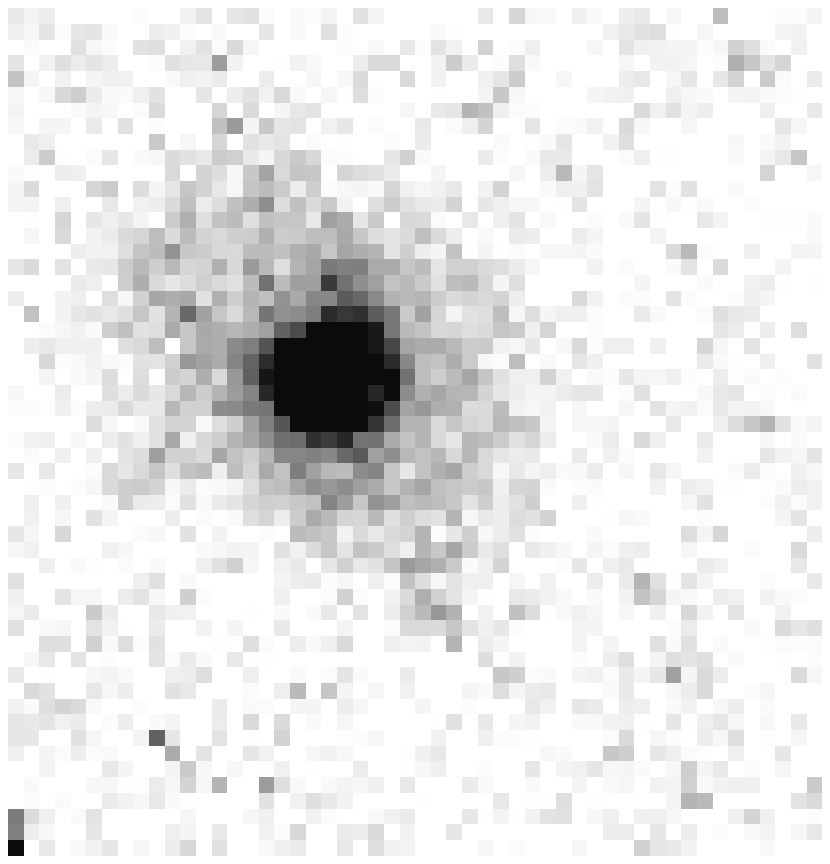}}
\mbox{\epsfysize=3cm \epsfbox{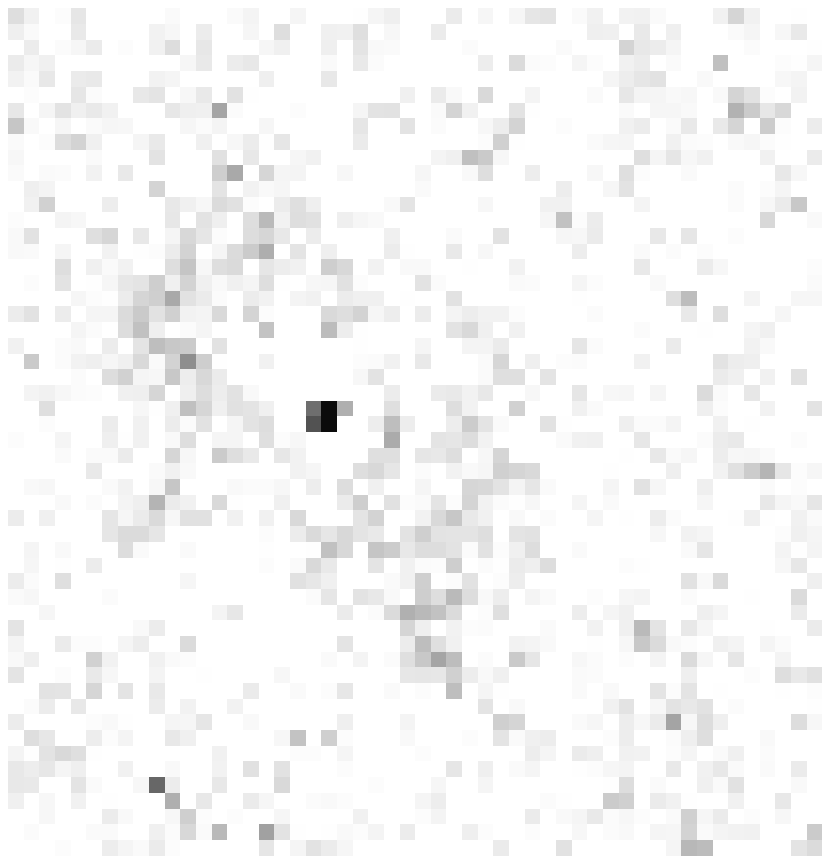}}
\mbox{\epsfysize=3cm \epsfbox{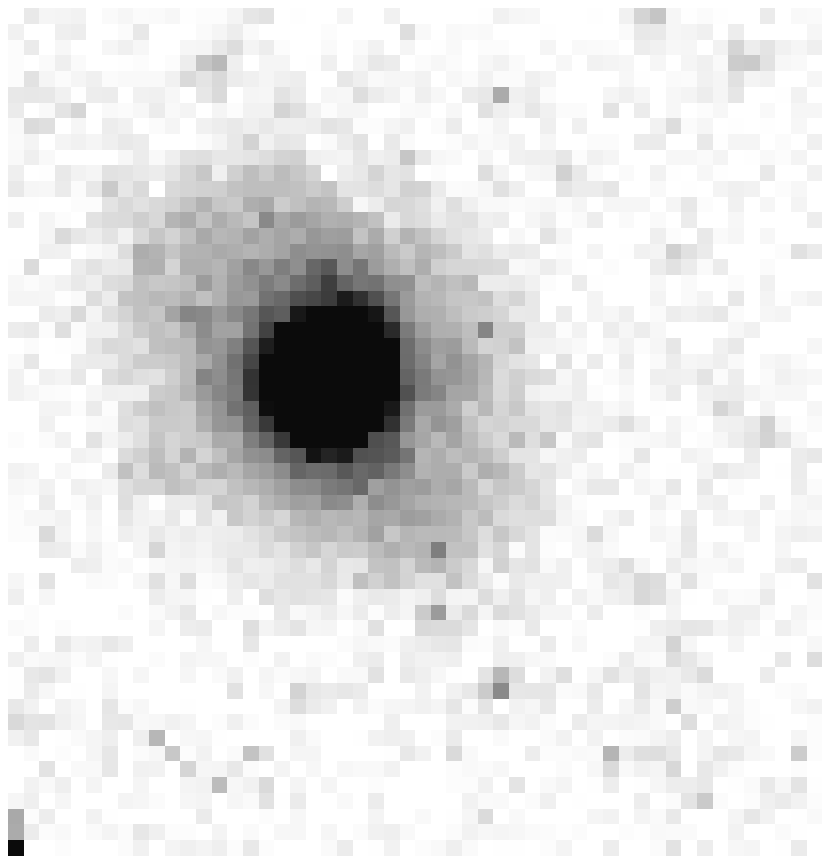}}
\mbox{\epsfysize=3cm \epsfbox{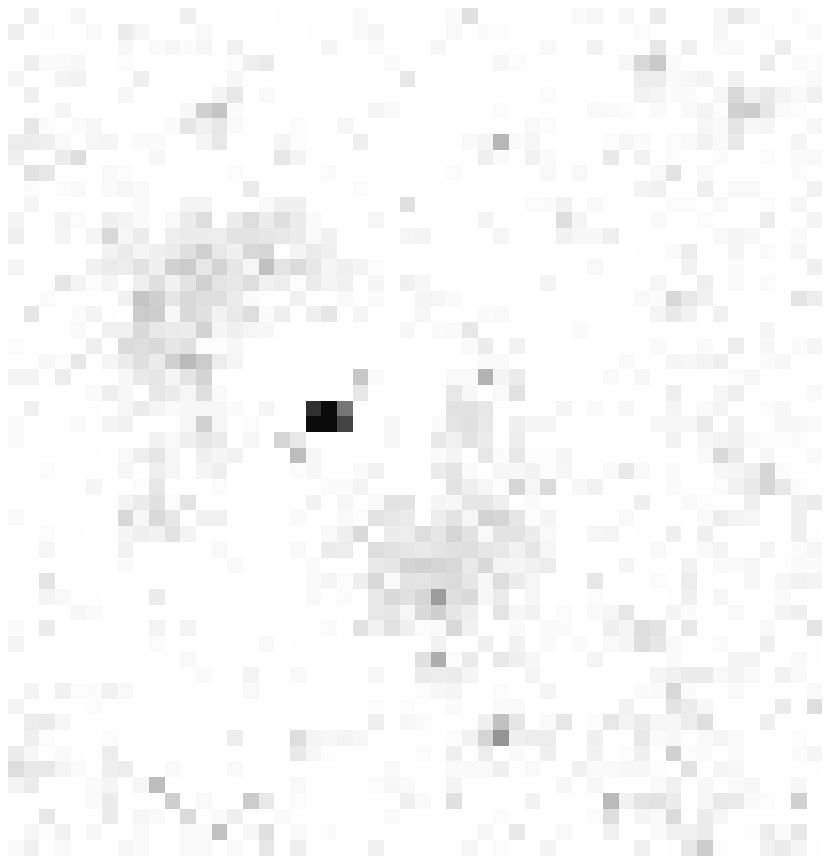}}}
\end{center}
\caption{WFPC2 images and 2D fit residuals for the galaxies without measured 
kinematics: {\bf B},{\bf O},{\bf P},{\bf Q}. In the residual of {\bf B} a
spiral pattern is recognizable.}
\label{fig:2Da}
\end{figure*}

For each galaxy we ran the code twice, minimising both the $\chi^2$
and the unweighted residuals (least squares). In the case of {\bf G}
the fit was not straightforward because of the companion {\bf H}.
Therefore we subtracted from the data, before performing the fit, a
model for {\bf H}, obtained from an \dv profile with $r_{\tx{e}}$,
SB$_{\tx{e}}$, $pa$, and $e$ taken from the isophotal profile fit.
The result was stable with respect to variations of the parameters of
{\bf H}.  In Figure~\ref{fig:2D} and \ref{fig:2Da} the original WFPC2
image and the fit residual are shown for all the galaxies fitted with
the 2D fits.

\subsubsection{Results}

All galaxies are well fitted by the \dv law (de Vaucouleurs 1948)
with the exception of galaxy {\bf B} which shows face-on spiral-like
residuals and might be an early-type spiral.

\begin{figure}
\mbox{\epsfysize=8cm \epsfbox{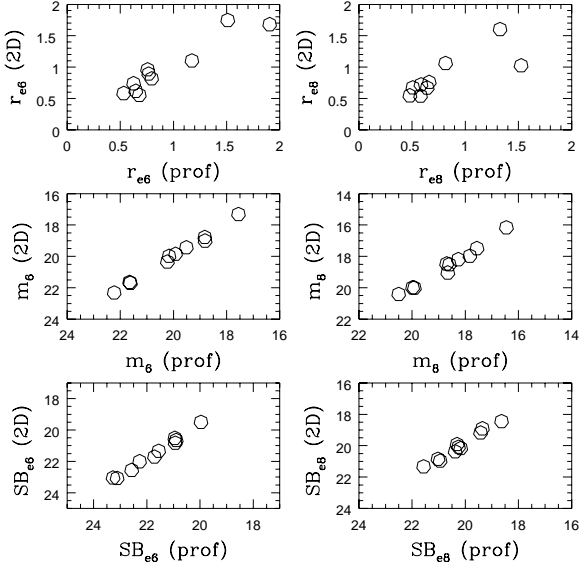}} 
\caption{Comparison of
 photometric parameters from the isophotal profile fit (x-axis) with
 those from one of the 2D fits (the least $\chi^2$, y-axis). Subscripts 6 and
 8 indicate filters F606W and F814W.} 
\label{fig:cf2D} 
\end{figure}

Standard magnitudes are obtained from the data following the Holtzman
et al.~\shortcite{WFPC2} calibration.  In Figure~\ref{fig:cf2D} we
plot the photometric parameters obtained from one of the 2D fits (the
least $\chi^2$) versus those given by the isophotal profile fit.  The
results are in very good mutual agreement, even though the procedures
differ in some important details, e.g., the isophotal profile fit is
in magnitudes and allows for $pa$ and $e$ variation, while the 2D fits
are in counts and have fixed $pa$ and $e$. As final value of the
parameters we use the average of the three results; the photometric
parameters are shown in Tables~\ref{tab:fotores} and
\ref{tab:fotores2}.  Integrated luminosities are those of the
best-fitting $r^{1/4}$ law.

\begin{table}
 \caption{Photometric parameters through filter F606W.  Errors (see
Section 3.4) do not include sky-subtraction contributions, which we
estimate to be 0.5 per cent on $r_{\tx{e}}$, 0.05 mag
$\tx{arcsec}^{-2}$ on SB$_{\tx{e}}$ and 0.02 mag on m (see text for
details).}
\label{tab:fotores}
\begin{tabular}{cllllll}
galaxy  & $r_{\tx{e}}$     & $\delta r_{\tx{e}}$ & SB$_{\tx{e}}$ &  $\delta \tx{SB}_{\tx{e}}$ & m  & $\delta \tx{m}$	\\
\hline
 A  &   1.65   &   0.13   & 20.64 &   0.13 &   17.57 &  0.12  \\
 B  &   0.588  &   0.038  & 19.73 &   0.11 &   18.90 &  0.05  \\
 D  &   0.793  &   0.042  & 20.99 &   0.09 &   19.5  &  0.03  \\
 E  &   0.794  &   0.071  & 21.60 &   0.13 &   20.12 &  0.06  \\
 F  &   1.145  &   0.017  & 22.56 &  $<0.01$ & 20.27 &  0.03  \\
 G  &   1.589  &   0.064  & 21.81 &   0.07 &   18.81 &  0.01  \\
 I  &   0.636  &   0.046  & 20.92 &   0.13 &   19.92 &  0.04  \\
 O  &   0.795  &   0.011  & 23.15 &   0.05 &   21.65 &  0.02  \\
 P  &   0.526  &   0.029  & 22.20 &   0.11 &   21.66 &  0.01  \\
 Q  &   0.612  &   0.018  & 23.20 &   0.06 &   22.27 &  0.02  \\
\hline
\end{tabular}
\end{table}
\begin{table}
\caption{Photometric parameters through filter F814W (same as Table 2).}
\label{tab:fotores2}
\begin{tabular}{cllllll}
galaxy  & $r_{\tx{e}}$ & $\delta r_{\tx{e}}$ & SB$_{\tx{e}}$ & $\delta \tx{SB}_{\tx{e}}$ & m & $\delta \tx{m}$\\
\hline
 A &  1.99  &  0.15  &	19.92 &  0.19 & 16.44 &  0.13 \\
 B &  0.548 &  0.014 &	18.58 &  0.05 & 17.89 &  0.04 \\
 D &  0.645 &  0.058 &	19.29 &  0.16 & 18.27 &  0.04 \\
 E &  0.872 &  0.080 &	20.31 &  0.12 & 18.64 &  0.07 \\
 F &  1.29  &  0.12  &	21.33 &  0.11 & 18.81 &  0.10 \\
 G &  1.419 &  0.075 &	20.29 &  0.09 & 17.53 &  0.02 \\
 I &  0.606 &  0.051 &	19.47 &  0.15 & 18.58 &  0.03 \\
 O &  0.654 &  0.007 &	21.02 &  0.05 & 19.94 &  0.02 \\
 P &  0.502 &  0.018 &	20.48 &  0.08 & 19.98 &  $<0.01$ \\
 Q &  0.547 &  0.053 &	21.14 &  0.17 & 20.48 &  0.03 \\
\hline
\end{tabular}
\end{table}

\subsection{A cross-check with the growth curve technique}

The growth curve technique is a standard procedure for nearby samples.
Two problems of this technique are: {\it i)} the statistical
dependence between different aperture magnitudes, e.g. in the presence
of light coming from a nearby galaxy or of diffuse background sources,
the systematic error affects the flux within all the apertures; {\it
ii)} the sensitivity to sky-subtraction errors.  We measured the
photometric parameters by fitting the growth curve of an $r^{1/4}$
profile to the aperture magnitudes and the results were in reasonably
good agreement with the other two methods.

\subsection{Error analysis}
\label{ssec:fotoerr}

We define as internal error of the parameters the standard deviation
of the results of the two 2D fits and the isophotal profile fit. Since
photon shot noise is negligible in our context, the main source of
error not included in the internal error is the uncertainty due to sky
subtraction (0.5 per cent on $r_{\tx{e}}$, 0.02 mag on m, 0.05 mag
$\tx{arcsec}^{-2}$ on SB$_{\tx{e}}$). We estimated this uncertainty by
measuring how the photometric parameters changed with the sky level,
and we added it quadratically to the internal error, so as to obtain
the total error; more precisely we used the variation of the
photometric parameters when the sky is shifted by one standard
deviation, calculated as the scatter of the sky levels measured in
different areas (a typical value is 2\%).  The photometric parameters
$r_{\tx{e}}$ and $ \langle I \rangle_{\tx{ e}} $ are correlated, as
can be seen from the errors calculated directly and reported in
Table~\ref{tab:fotcorr}.  The particular combinations of effective
radius and surface brightness are chosen because they enter the
definition of the $\kappa$ space (Bender, Burstein \& Faber 1992).
Also the error on the colour ($m_{\tx{\small F606W}}-m_{\tx{\small
F814W}}$) is different from the sum of errors on the single filter
magnitudes (see Table~\ref{tab:fotcorr}) since the errors in the two
filters are correlated: it generally happens that when one method
(e.g. 2D $\chi^2$) gives a lower magnitude than another (e.g. profile
fits), the same thing is repeated in both filters.

\begin{table}
\caption{Photometric parameters II. Errors are shown as examples of
the correlation of the photometric uncertainties. The quantity
$colour$ is defined as $m_{\tx{\small F606W}}-m_{\tx{\small F814W}}$.}
\begin{tabular}{ccccc}
galaxy & colour & $\delta$colour & $\delta \langle I \rangle_{\tx{e}}^2
r_{\tx{e}}^{-1}/ \langle I \rangle_{\tx{e}}^2 r_{\tx{e}}^{-1}$ & $\delta \langle I \rangle_{\tx{e}} r_{\tx{e}}/ \langle I \rangle_{\tx{e}} r_{\tx{e}}$ \\ 
\hline 
A & 1.127 & 0.012 & 0.23 & 0.089 \\ 
B & 1.015 & 0.012 & 0.24 & 0.047 \\ 
D & 1.226 & 0.018 & 0.19 & 0.027 \\ 
E & 1.486 & 0.013 & 0.29 & 0.034 \\ 
F & 1.463 & 0.068 & 0.025 & 0.011 \\ 
G & 1.273 & 0.007 & 0.15 & 0.028 \\ 
I & 1.333 & 0.005 & 0.28 & 0.044 \\ 
O & 1.705 & 0.006 & 0.091 & 0.028 \\ 
P & 1.678 & 0.013 & 0.26 & 0.045 \\ 
Q & 1.789 & 0.051 & 0.15 & 0.031 \\
\hline
\end{tabular}

\label{tab:fotcorr}
\end{table}

\section{Spectroscopy}
\label{sec:spec}

The spectroscopic observing run took place in the period April 18-22
1996 at the 3.6-m ESO telescope.  Long slit ($ 1 \farcs 5 $ slit width)
spectroscopy was obtained with the EFOSC spectrograph, with the Orange
150 grism in the spectral range between approximately 5200 and 7000
\AA\, (512$\times$512 CCD, with pixel size 30 $\mu m$ equivalent to
about $ 0 \farcs 57 $).  The slit length of about $ 180 \arcsec $
projects to only about 320 pixels on the CCD.  For each target we
obtained multiple exposures.  The estimated seeing, measured with R-band
acquisition images, was $ 0 \farcs 8 $ FWHM.  Table~\ref{tab:splog}
gives the basic observation log. 

\begin{table}
\caption{Spectroscopic data: number of observations (nexp), total exposures
times (texp), and average signal-to-noise ratio per pixel (S/N).}
\label{tab:splog}
\begin{tabular}{|l|r|r|r|}
galaxy & nexp & texp (s) & S/N \\
\hline
 A  &  4   &  7200  &  65 \\
 B  &  4   &  7200  &  40 \\
 D  &  4   & 12000  &  22 \\
 E  &  5   & 13200  &  17 \\
 F  &  5   & 13200  &  12 \\
 G  &  3   & 10800  &  35 \\
 I  &  3   & 10800  &  22 \\
 LM &  2   &  4800  &  - \\
 N  &  2   &  4800  &  - \\
 O  & 10   & 36000  &  7 \\
 P  & 10   & 36000  &  6 \\
 Q  & 10   & 36000  &  - \\
\hline
\end{tabular}
\end{table}

The slit positions were carefully chosen to include as many objects
(two or three at a time) as possible.  Every night we obtained an
exposure with the He-Ar lamp at zenith; before and after every galaxy
spectrum we took a He lamp spectrum as a check of the dispersion
relation shifts.  We observed two spectrophotometric standard white
dwarfs in order to perform relative-flux calibration of the spectra.
The identification of our targets as normal, non star-forming,  
ellipticals was confirmed by the absence of emission lines in their
spectra.

\subsection{Reduction}

A superbias file was obtained by averaging about a hundred bias
images and was subtracted from the data. Dome flatfield exposures
were used to produce a flatfield along the wavelength direction (CCD
columns). The slit function was obtained from sky frames and was
found to be independent of wavelength. Dark current subtraction was
done by producing a dark frame from the average of 8 darks 1 hour long. Since this mean dark did not contain structures, we only
subtracted the mean value.

Throughout the reduction and analysis we have avoided (non-linear)
rebinnings. To this end, we did the reduction before wavelength
calibration. The extraction of the spectra was a crucial point of the
reduction process, given the faintness of the signals.  Averaging too
many rows would have meant too much noise, using only the central row
would have meant losing important signal.  As we could not simply
extract a fixed pixel width (along the spatial direction) because of
optical distorsions, we used a code developed by M{\o}ller \&
Kj{\ae}rgaard (1992) to obtain a one-dimensional spectrum, extracted
with optimal signal-to-noise weighting. The code also removes cosmic
rays, identified as pixels five standard deviations away from the
expected signal.  All spectra were carefully inspected by eye before
and after extraction to check the proper cosmic ray removal.

\begin{figure*}
\mbox{\epsfysize=18cm \epsfbox{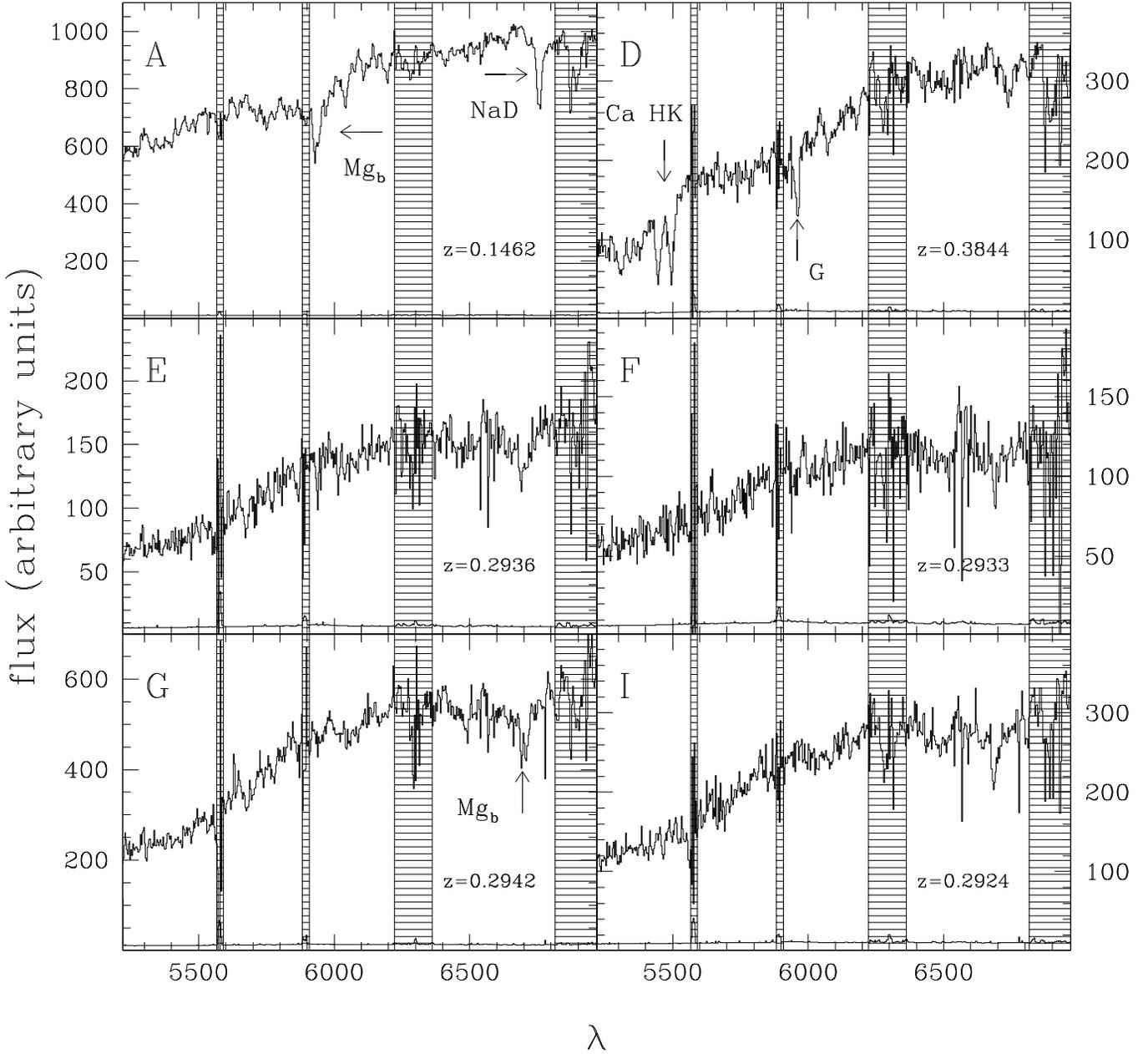}}
\caption{Wavelength (in \AA) calibrated spectra of the galaxies with
measured velocity dispersion. The shaded spectral regions were
affected by strong sky-emission lines. These regions have not been
used in the kinematic fit. The standard deviation per pixel is also
shown (lower curves). The arrows identify some of the main absorption
features, such as Ca H (3968.5 \AA) and K (3933.7 \AA), the G band
(4304.4 \AA), Mg$_b$ (5175.4 \AA) and Na D (5892.5 \AA).}
\label{fig:spectra}
\end{figure*}

Given the faint signal of the spectra, we took special care during sky
subtraction.  For each galaxy we extracted two sets of columns, to the
left and to the right of the galaxy.  The columns were selected to be
as close as possible to the galaxy to limit optical distorsion
problems.  Columns of one set were combined neglecting the highest and
lowest pixels to remove cosmic rays.  The result was eye-inspected and
then the two spectra obtained were averaged to obtain a high
signal-to-noise sky spectrum.  This procedure allowed us to check the
optical distorsion by comparing the emission lines between the left
and the right spectra.  The sky spectra obtained had broader lines
than the single column ones, because we combined spectra
wavelength-shifted by optical distorsion.  Sky contribution was
subtracted during spectrum extraction by deconvolving the artificial
additional width.  The resulting one-dimensional spectra were combined
to obtain the final galactic spectrum, after verifying that the
relative shifts of the positions along the y-axis of the emission
lines were negligible. The reduced spectra are shown in Figure
\ref{fig:spectra}. The average signal-to-noise ratio per pixel of the
spectra is listed in Table \ref{tab:splog}.  The average
signal-to-noise ratio ranges from 65 per pixel for the brightest
galaxy ({\bf A}) to 12 per pixel for the faintest galaxy for which we
obtained a velocity dispersion ({\bf F}); the estimated error is
accordingly larger for {\bf F} than for {\bf A}. For comparison,
Davies et al. (1987) in their low redshift study, with a total
signal-to-noise ratio equivalent to 5$\cdot10^5$ photons, estimate the
random error to be about 10 \%.  The analysis in Section \ref{sec:res}
is done both with the entire sample and excluding the two galaxies
{\bf E} and {\bf F} with the lowest signal-to-noise spectra among
those with measured velocity dispersion.

\subsection{Wavelength calibration and measurement of the instrumental
resolution}

\label{sec:wl}

We found the dispersion relation for each He-Ar lamp using the {\sc
midas gui/long} software.  The different calibrations were all within
the errors.  We used short He lamp images to check that shifts between
single exposures were negligible.  The He-Ar lamps were also used to
measure the instrumental resolution: the line shapes were fitted with
Gaussian functions to obtain line widths as a function of CCD pixel
number, and, via dispersion relation, of wavelength.  The FWHM is independent of $\lambda$ within the measurement scatter and it was determined to be: 
\begin{equation} \Delta
\lambda = (8.39\pm 0.26) \tx{\AA}, 
\label{res}
\end{equation}
Translated into velocities it corresponds to a kinematic resolution
ranging from 206 to 155 \kms\, from the blue to the red. The 3 \%
error is due to the scatter in the measurement (see also Section
4.4). The instrumental resolution measurement is very important for the
kinematic fit, as we discuss in Section 4.3.  Therefore we carefully
examined all lines, to avoid ghost lines which were found and which
would have affected the measurement. We checked the resolution by
measuring the width of the sky lines and the results were consistent
with the ones found with the He-Ar lamp (8.11$\pm$0.39 \AA).

For comparison, the instrumental resolution of the Lick/IDS
spectrograph was between 8-10 \AA\, at shorter wavelengths (see
e.g. Dalle Ore et al., 1991). The velocity dispersions we find are
larger than half the instrumental resolution, which is commonly
considered as a limit (see e.g. Dressler 1979; Kormendy 1982; Bender,
1990). It has been noticed that when the velocity dispersion becomes
of the order of the kinematic resolution and the signal-to-noise is
low a systematic error may be introduced (see e.g. J\o rgensen, Franx
\& Kj\ae rgaard, 1995). In order to study this effect we broadened
some stellar templates to 150,~200,~250,~300 \kms and we added noise, thus
producing artificial galaxies with S/N per pixel ranging between 10
and 60. The artificial galaxies were then fitted with the same stellar
templates to measure how well the original velocity dispersions were
recovered.  The largest velocity dispersions were recovered within few
percents (2--4~\%) down to S/N=10, while the smaller one (150 \kms)
resulted systematically underestimated by $\sim10$\% at S/N 10. No
systematic trends were noticed above S/N=15. This bias is negligible
to our measurement, as galaxies {\bf E} and {\bf I}, the only with
measured velocity dispersion below 200 \kms (see Table 7), have high
enough S/N (see Table \ref{tab:splog}), while {\bf F}, the only with
S/N below 15 has $\sigma\approx 200$ \kms.

\subsection{Kinematics}
\label{ssec:km}

In order to derive a robust measurement of the kinematics of our
sample of galaxies we used two different and independent codes: {\it
i)} the {\sc Gauss-Hermite Fourier Fitting Software} \cite{gh1,gh2},
hereafter GHFF, and {\it ii)} the {\sc Fourier Quotient} \cite{fq},
hereafter FQ, in the version modified by Rose (Dressler, 1979) and
used e.g. by Stiavelli, M{\o}ller \& Zeilinger (1993). The results of
the two methods are mutually consistent, within the estimated
errors. However, the GHFF, allows a reliable estimate of the errors
(see, e.g., Rix \& White, 1992, and our discussion below) while the
version of the FQ we used does not; moreover methods like the GHFF are
known to be less sensitive to template mismatches than the FQ (see
e.g. Bender 1990, and our discussion below). Furthermore the GHFF
provides more insight in the fitting procedure, because the continuum
subtraction, the tapering, and the filtering of both observed and
template spectrum can be checked step by step. Finally it provides a
quality parameter (the $\chi^2$) and the residual spectrum.  Therefore
we shall consider the values obtained with the FQ only as a check of
the values obtained with the GHFF. We report the complete results
obtained with both codes (Tables 7 and 8 for the GHFF, Table 9 for the
FQ). The discussion of the FP properties in Sections \ref{sec:res}
and \ref{sec:conc} is based on the GHFF method, but the FQ method
leads to the same conclusions.

\subsubsection{Template spectrum}

We used star spectra from the Jacoby et al.~\shortcite{J84} library as
starting bricks for the template construction.  We simulated the
measured experimental resolution, taking into account the
non-negligible width of the original Jacoby spectra (4.5 \AA\,
FWHM), by convolving the redshifted template spectra
with a gaussian of width $\sigma_{t}^2=8.39^2-(4.5(1+$z$))^2$, where
$z$ is the galaxy redshift\footnote{The redshift was first determined
by recognizing the most evident absorption lines and then by iteratively refining the
measurement.}.  The choice of the correct spectral
class for the template is critical in order to avoid systematic
errors.  Therefore, we decided to fit the galactic spectra with each
of the spectral types from F0 to K7 to be sure to include all the most
relevant spectral types. By setting some fit quality criteria (see the
following section) we selected a set of ``good'' spectral types. The
range of variation of the central velocity dispersion within the
``good'' spectral types provides a check of the robustness of the
result, and at the same time an estimate of the systematic error
possibly induced by residual template mismatches.  We avoided the
alternate approach of using synthetic galaxy spectra because of its
heavy model dependence on the underlying stellar population.

\subsubsection{The {\sc Gauss-Hermite Fourier Fitting Software}(GHFF) fit}

The GHFF, developed by R.~P. van der Marel \& M.~Franx (Franx,
Illingworth \& Heckman, 1989; van der Marel \& Franx, 1993) models the
broadening function $B$ as the first terms of a Gauss-Hermite series,
thus allowing one to derive a better estimate of the velocity
dispersion in case of non-Gaussian broadening. The advantages and
caveats of this approach are discussed in detail in van der Marel \&
Franx (1993). In order to achieve a meaningful estimate of the first
non-Gaussian terms however, very high signal-to-noise ratio is
required \cite{gh1,fcq} and generally the studies on the Fundamental
Plane with this software fit only the Gaussian component (see e.g. J\o
rgensen, Franx \& Kj\ae rgaard, 1995). Therefore we used the code in
Gaussian-fitting mode. 

The GHFF code provides the redshift (\zgh), the line strength ratio (\ggh) and the velocity dispersion (\sgh). The fit gives also an error value
for the parameters that we will call $\delta v_{\tx{ghff}}$, $\delta
$\ggh, $\delta \sigma_{\tx{ghff}}$.

Generally, the templates within the spectral types G0-K2 provided the
best fit, as could be judged by looking at the $\chi^2$, the residuals
and the broadened template, but no template was clearly a better
approximation than the rest.  In order to better understand the
systematic trends, we created some ``artificial galaxies'' by
broadening the templates and adding noise to reproduce a
signal-to-noise ratio of about 12 per pixel. The GHFF with the
templates in the range G0-K2 was able to recover the correct velocity
dispersion, while generally the F stars gave lower sigmas and the K3-7
higher ones. The scatter in the values obtained with different noise
patterns was comparable with the formal error of the fit when the same
star was used both as ``galaxy'' and template, but was higher even in
case of small mismatching. For example a ``galaxy'' produced and
fitted with a G5 spectrum gave on average 96~\% the correct value,
with 15~\% average formal error and a measured scatter of 17~\%; using
a G2 star as template we obtained equally 96~\% of the correct value,
with 22~\% scatter and 17~\% average formal error; with a G6 the same
figures were 104~\%, 21~\% and 19~\%.  Therefore we decided to keep
only the G0-K2 interval, where no systematic trend was noticed.

We defined the scatter of the velocity dispersion of the ``good''
templates to be the estimate of the systematic error due to residual
mismatches. In order to obtain results comparable to the FQ ones, we
selected the good fits with criteria applicable also to the FQ (that
in the version we have does not provide the $\chi^2$). We used the
information provided by the code as an indicator of the quality of the
fit. Specifically, we used the line strength ratio \ggh as an
estimator of the mean ratio of the abundances of the absorbers (i.e.,
roughly speaking, the metallicity) and $\delta$\sgh as a measure of
the mismatch in the shape (primarily due to differences in the
spectral type).  A fit is considered acceptable if it satisfies the
three conditions: {\it i)} 0.5 $<$ \ggh $<$ 1.5, {\it ii)}
$\sigma_{\tx{gh}}>0$ and {\it iii)} $\delta$\sgh$<$\sgh. As final
value we adopt the average of all acceptable results weighted on the
fit formal errors.

We checked what templates were rejected by
this ``$\chi^2$ blind'' algorithm, and the criteria were very
effective in selecting the best $\chi^2$ and the best matching
templates. The GHFF compared to the FQ produced generally a larger
number of fits within the quality criteria and a lower scatter as a
function of template, thus confirming to be less sensitive to template
mismatches.  As a further check we also considered a simple average of
the five best $\chi^2$ and the changes were much smaller than the
error bars.  The average values are listed in Table
\ref{tab:cineresGH}. The fit formal errors listed refer to the
template which gave the least $\chi^2$. 

We have omitted galaxy {\bf B} in our kinematic discussion since its
kinematic fit does not satisfy our quality criteria and indicates a
very small velocity dispersion, below our resolution limit and
consistent with the morphological appearance as a face-on early spiral
(see Figure~\ref{fig:2Da}).

\begin{table}

\caption{Kinematic results from the GHFF fit I. For each galaxy we list
the average central velocity dispersion (\sgh) together with the
random error ($\delta$ \sgh) and the systematic error ($\Delta$
\sgh). The random error is the fit formal uncertainty for the
best-fitting template, while $\Delta$ \sgh is the scatter of the
values obtained with the ``good'' templates (see text).}
 
\label{tab:cineresGH}
\begin{tabular}{lccc}

gal & $\sigma_{\tx{ghff}}$    & $\delta\sigma_{\tx{ghff}}$ &
$\Delta$\sgh \\

\hline
  A     &  232 	&  11  & 28  \\
  D     &  271  &  24  & 16  \\
  E     &  175  &  28  & 28  \\
  F     &  195  &  41  & 74  \\
  G     &  226  &  15  & 25  \\
  I     &  166  &  21  & 29  \\
\hline
\end{tabular}
\end{table}

\begin{table}

\caption{Kinematic results from the GHFF fit II. Same as Table
\ref{tab:cineresGH} for the redshifts (\zgh) and line strength ratios
(\ggh).}
 
\label{tab:cineres2GH}
\begin{tabular}{lcccccc}

gal & \zgh & $\delta$ \zgh & $\Delta$ \zgh & \ggh & $\delta$ \ggh & $\Delta$ \ggh\\
\hline
  A &  0.1462 & 3$\cdot 10^{-5}$& 3$\cdot 10^{-4}$ &  0.90  &   0.02 & 0.14\\
  D &  0.3844 & 5$\cdot 10^{-5}$& 1$\cdot 10^{-4}$ &  0.80  &   0.03 & 0.04\\
  E &  0.2936 & 6$\cdot 10^{-5}$& 1$\cdot 10^{-4}$ &  0.91  &   0.04 & 0.14\\ 
  F &  0.2933 & 1$\cdot 10^{-4}$& 1$\cdot 10^{-4}$ &  0.95  &   0.10 & 0.19\\
  G &  0.2942 & 4$\cdot 10^{-5}$& 1$\cdot 10^{-4}$ &  0.93  &   0.03 & 0.16\\  
  I &  0.2924 & 4$\cdot 10^{-5}$& 1$\cdot 10^{-4}$ &  0.84  &   0.04 & 0.16\\  
\hline
\end{tabular}
\end{table}

\subsubsection{{\sc Fourier Quotient}(FQ) fit}
\label{sssec:fq}

The basic idea of the FQ is the hypothesis that one can
describe a galaxy spectrum  G as the convolution $\tx{G}=\tx{B}\circ
\tx{S}$ of an average star template spectrum S with a broadening
function B.  The broadening function is purely kinematic, while the
template spectrum includes the broadening due to the instrumental
resolution.  The template spectrum has to be a good approximation of
the galaxy spectrum especially in its absorption features because FQ
works on the spectra after subtraction of the continuum.  Selecting
the best template is critical to the measurement since, as shown,
e.g., by Bender \shortcite{fqc}, FQ is very sensitive to the spectral
type of the star used as a template. To perform the kinematic fit with
the FQ we needed spectra free of their continuum component.  The
continuum was subtracted out by interpolating between regions with no
spectral features.

The version of the FQ code available to us requires a
template and a galaxy spectrum linearly rebinned in wavelength and
then it rebins them in $\log \lambda$.  We modified it to avoid the
extra rebinning. The new version requires the one-dimensional galaxy
spectrum as a function of the CCD pixel number and the dispersion
relation given as polynomial coefficients; the rebinning is done
directly in $\log{\lambda}$. The spectral regions affected by the
strongest sky emission lines were masked out.

Free parameters in the fit are the redshift
(\zfq), the width (\sfq), and a normalising
factor (\gfq, the line strength).  The fit gives also a formal error value
for the parameters that we will call $\delta v_{\tx{fq}}$, $\delta
$\gfq, $\delta \sigma_{\tx{fq}}$.

A detailed study of the results shows no correlations between fit
parameters and errors.  The only general trend is that best results
are obtained for templates of spectral type G, and they become worse
towards the edges of the range we considered, F0 and K7 (for the
complete results see Treu 1997), consistently with the findings of the GHFF. Generally, the F star templates gave
systematically lower velocity dispersion, while the K3-7 templates
gave systematically higher values (see also the discussion in the GHFF
section). In Table
\ref{tab:cineres} we list the average velocity dispersion, line
strength ratio and redshift, together with the scatter in the results between different templates as for the GHFF.

Consistently with what is
found with the GHFF, the velocity dispersion derived for {\bf B} was
below our resolution limit.

\begin{table}
\caption{Kinematic results from the FQ fit. For each galaxy we list
the average central velocity dispersion, redshift and line strength ratio (\sfq,\zfq, \gfq) together with the
systematic error ($\Delta$ \sfq, $\Delta$ \zfq, $\Delta$ \gfq), that is the scatter of the values
obtained with the ``good'' templates (see text).}
 
\label{tab:cineres}
\begin{tabular}{lcccccc}

gal & $\sigma_{\tx{fq}}$ & $\Delta$\sfq & $z_{\tx{fq}}$ & $\Delta$\zfq & \gfq  & $\Delta$\gfq\\
\hline
  A	&     208 & 39  & 0.1462 & 1$\cdot 10^{-4}$ & 0.92 & 0.16\\
  D	&     252 & 13  & 0.3844 & 1$\cdot 10^{-4}$ & 0.84 & 0.14\\
  E	&     211 & 10  & 0.2935 & 2$\cdot 10^{-4}$ & 0.67 & 0.10\\
  F	&     233 & 74  & 0.2934 & 2$\cdot 10^{-4}$ & 0.61 & 0.01\\
  G	&     216 & 55  & 0.2924 & 1$\cdot 10^{-4}$ & 0.56 & 0.03\\
  I	&     214 & 39  & 0.2918 & 2$\cdot 10^{-4}$ & 0.65 & 0.10\\
\hline
\end{tabular}
\end{table}

\subsection{Error analysis}
\label{sec:cinerr}

The errors on $\sigma$, $\gamma$, $z$ are a combination of a
systematic component due to template errors and mismatch, and a random
component due to the limited signal. For the systematic error we adopt
the standard deviation $\Delta \sigma$ of the sample of the acceptable
$\sigma$ values (see Tables~\ref{tab:cineresGH} and
\ref{tab:cineres}). For the random error, we adopt the formal error
given by the GHFF fit with the best-fitting template, although it is
probably an overestimate because it includes possible residual
template mismatches.  The random and systematic errors for the line
strength ratio ($\gamma$) and the redshift ($z$), defined in the same
way, are listed in Tables \ref{tab:cineres} (FQ) and
\ref{tab:cineres2GH} (GHFF).

The uncertainty in the measurement of the instrumental resolution
introduces a systematic error in the measurement of the velocity
dispersion that is in our case negligible. For example, considering
{\bf A}, the galaxy with the highest signal-to-noise spectrum, the
random error is estimated by us to be 5 \%, the systematics due to
template mismatches 12 \%, and therefore a further 3 \%, to be added
in quadrature, does not change the total error significantly.  As a
double check we repeated the fits with templates with different
resolution ($\delta \sigma_t$ equivalent to 0.26 \AA) and the results
changed as expected:
\begin{equation}
\label{eq:resch}
\delta \sigma = \delta \sigma_t \left(\frac{\sigma_t}{\sigma}\right).
\end{equation}
For galaxy {\bf I} we obtained the larger difference (7 \kms), still
negligible with respect to the other sources of error.

The spectrum of galaxy {\bf D} includes the lines Ca H and K. While
widely used to perform such measurements (e.g. Dressler 1979), these
lines has been reported to induce a slight overestimate of the
velocity dispersion (Kormendy, 1982, estimates it to be between 5 and
20 \%; see also Kormendy and Illingworth, 1982). They suggest the
problem may be caused by the intrinsic width of the lines or by the
steepness of the continuum in that spectral region. However, Dressler
(1984), suggest that they may be best suited for measurement of large
velocity dispersion for faint objects, such as {\bf D}. For these
reasons we performed the kinematic fit also exluding the region of Ca
H and K and we found a velocity dispersion 10 \% higher in this case,
a variation of the order of the estimated errors.  In the analysis presented in this paper we shall
consider the results from the entire spectrum, for its higher
signal-to-noise ratio. Nevertheless, when a larger sample of these
objects will be available we plan to better address this issue by
studying systematic variations induced by Ca H and K on the sample.

Similarly the spectrum of {\bf A} includes the region of NaD, which
may be affected by interstellar absorption even in ellpitical
galaxies (e.g. Dressler, 1984). We repeated the analysis exluding the
NaD region and found that this resulted in a 3\% increase in the velocity dispersion, again within
the estimated errors.

The error on $z$ is caused by uncertainties in the wavelength
calibration (the dominant source), in the fit results and in the
absolute motion of the template stars, for the Jacoby et
al.~\shortcite{J84} spectra were not corrected for peculiar motions
(estimated to be of order 0.0001 by the authors).  The total error
adds up to about 0.0005.  In the cases with insufficient signal, the
redshift has been measured by identifying the main spectral features,
with a total estimated error of $\approx0.005$.  These redshifts were
then been checked via superimposition of redshifted templates. The
listed error corresponds to $\sim 10$ pixel in the studied spectral
range.

\subsection{Relative flux calibration}

The instrumental throughput is not constant as a function of
wavelength.  In order to obtain a good relative flux calibration for
our spectra, we observed two spectrophotometric standard white dwarfs
through our instrumental setup for a total of five exposures.  By
comparison with the spectra taken from the {\sc midas} libraries, we
obtained the response function separately for each exposure, using
{\sc midas gui/long} and the values of galactic extinction from
Burstein \& Heiles \shortcite{maps}.  The results were all within two
percent and we adopted their average function to correct the
spectra for the instrumental sensitivity.

\section{Photometric and kinematic corrections}
\label{sec:corr}

The spectroscopic and photometric results described so far require a number of
corrections.  Photometry should be corrected for galactic extinction
(Subsection~\ref{ssec:gaex}), while WFPC2 magnitudes should be
transformed to standard restframe magnitudes (Subsection~\ref{ssec:K})
and $\sigma_{\tx{ghff}}$ should be corrected into a standard central
velocity dispersion $\sigma$ (Subsection~\ref{ssec:scorr}). 

\subsection{Galactic extinction}
\label{ssec:gaex}

For galactic extinction, we used E(B-V) values from the Burstein \&
Heiles \shortcite{maps} maps, and the relations $A_{\tx{\small F606W}}=2.907\,
\tx{E(B-V)}$ and $A_{\tx{\small F814W}}=1.902\, \tx{E(B-V)}$, calculated by
Romaniello \shortcite{Rom}.  The extinction coefficients obtained for
the various fields are given in Table~\ref{tab:gaex}.

\begin{table}
\caption{Galactic extinction coefficients.}
\label{tab:gaex}
\begin{tabular}{lccc}
 field	& E(B-V)		& $A_{\tx{F606W}}$	& $A_{\tx{F814W}}$\\
\hline		
 u5405 	& 0.18-0.21 & 0.57 & 0.37 \\	 
 ur610	& 0.03-0.06 & 0.13 & 0.09 \\
 urz00	& 0.06-0.09 & 0.22 & 0.14 \\
 ust00	& 0         & 0	   & 0    \\	
 ut800	& 0.03-0.06 & 0.13 & 0.09 \\
\hline
\end{tabular}
\end{table}

\subsection{An improved way to calculate K-correction}
\label{ssec:K}

The photometric parameters measured through WFPC2 filters have different 
physical meaning depending on the galaxy redshift. The common technique to 
obtain standard magnitudes is to introduce the so-called K-correction,
 \begin{equation}
  \textrm{K}\equiv-2.5\log\left[\frac{\int F_{\lambda}
     (\lambda)S_o(\lambda)d\lambda}{\int F_{\lambda} (\lambda)S_o
     (\lambda(1+z))d\lambda}\right],
 \end{equation}
 where $F_{\lambda}$ is the spectral flux density of the galaxy and
$S_{o}$ is the response function of the instrumental setup used during the
observations.  To calculate the K-correction at intermediate redshift,
knowledge of a large spectral range is required.  In order to reduce the
required spectral range and to minimise the correction, it is convenient
to calculate the restframe absolute magnitudes ($M_{\tx{w}}$) in different
filters from the observational ones ($m_{\tx{o}}$):
 \begin{equation}
  M_{\tx{w}}\equiv \Delta m_{\tx{wo}}(z;F_{\lambda}) + m_{\tx{o}} - DM(z;\Omega,\Omega_{\Lambda}),
 \end{equation}
where
\begin{eqnarray}
  \Delta m_{\tx{wo}}\equiv&-2.5\log\Big[\frac{\int F_{\lambda}(\lambda)S_{\tx{w}}(\lambda)d\lambda \nonumber}{\int F_{\lambda}^{st}(\lambda)S_{\tx{w}}(\lambda)d\lambda}\Big]\nonumber\\ 
&+2.5\log\Big[\frac{\int F_{\lambda}(\lambda)S_{\tx{o}}(\lambda (1+z))d\lambda}{\int F_{\lambda}^{st}(\lambda)S_{\tx{o}}(\lambda)d \lambda} \Big],
\end{eqnarray}
$DM(z;\Omega,\Omega_{\Lambda})$ is the bolometric distance modulus,
and $F_{\lambda}^{st}$ is the spectral flux density of a standard
star.  For each redshift, we calculated corrections from F606W to
Johnson B ($\Delta m_{\tx{\small BF606W}}$) and from F814W to Johnson
V ($\Delta m_{\tx{\small VF814W}}$).

To perform this calculation we chose the following, purely empirical,
approach. For each galaxy we selected from the synthetic library of
Bruzual \& Charlot (1993; GISSEL96 version) the spectrum which best
approximates the galactic spectrum under consideration in the observed
spectral range. Assuming that the selected spectrum provides a good
description of the spectral energy distribution of our galaxy also
outside the observed wavelength range, we used it to compute $\Delta
m_{\tx{wo}}$.  This method also provided us with an error estimate based on
the range $\Delta m_{\tx{wo}}$ obtained from all the spectra not
significantly different from the observed one; the typical error on
the K-correction is $\approx 0.05$ mag.  Since we were particularly
interested in the continuum shape, which is the main contribution to
large-passband fluxes, we needed flux calibrated spectra.  The
so-called age-metallicity degeneracy \cite{Worthey} with respect to
the continuum shape is not a significant concern, because we were only choosing
the best phenomenological approximation.  In Table~\ref{tab:Kcorr} we
list the computed $\Delta m_{\tx{wo}}$ values.  As a cross-check we
compared the WFPC2 colours of our galaxies with the WFPC2 colours
``measured'' on the redshifted synthetic spectra.  The colours from
the best approximation spectrum were all within 0.05 magnitudes from
the observed one.

\begin{table}
\caption{Photometric corrections.}
\label{tab:Kcorr}
\begin{tabular}{lccc}
galaxy& $z$  &	$\Delta m$ & $\Delta m$\\
& 	& B F606W &	V F814W	 \\
\hline
A  & 0.146 &   1.11	& 1.12					    \\
B  & 0.146 &   1.01	& 1.04					    \\
D  & 0.384 &   0.44  	& 0.86					    \\
E  & 0.294 &   0.68     & 1.00					    \\
F  & 0.293 &   0.68	& 0.99					    \\
G  & 0.294 &   0.63	& 0.92					    \\
I  & 0.292 &   0.65	& 0.94					    \\
O  & 0.647 &   -0.28	& 0.65					    \\
P  & 0.550 &   -0.05	& 0.74					    \\
\hline	       					
\end{tabular}

\medskip

For the galaxies {\bf B}, {\bf O}, {\bf P} with insufficent
signal-to-noise ratio, the redshift was found by identifying the
strongest spectral features. The redshift of {\bf P} is somewhat less
certain because the signal-to-noise ratio is particularly low.

\end{table}

\subsection{Reduction to a standard central velocity dispersion}
\label{ssec:scorr}

In Section~\ref{sssec:fq} \sgh was defined as the broadening of the
projected spectrum integrated over the solid angle subtended by the
CCD.  The velocity dispersion in elliptical galaxies varies with
distance from the center.  Usually, the velocity dispersion is defined
as the kinematic broadening of a spectrum integrated over a reference
solid angle.  The choice of a standard reference solid angle makes it
possible to compare values obtained in different studies.  Previously
used reference solid angles are, e. g., a fixed angular size at a fixed
distance or a fraction of the effective radius (J{\o}rgensen, Franx \&
Kj{\ae}rgaard 1996; hereafter JFK96).  The latter choice is to be
preferred because it is related to an intrinsic length scale of the
galaxy and does not depend on cosmological parameters. Following
JFK96, we correct all measurements to a circular aperture of $r_{\tx{e}}/4$.
To calculate the correction we need to model the large-scale kinematic
profile of elliptical galaxies. A power law:
\begin{equation} 
\sigma(r)\propto \left(\frac{r}{r_e}\right)^d,
\end{equation} 
is generally a good description of the kinematic
profile (see, e.g., the radially extended profiles in Carollo \&
Danziger 1994a, 1994b, Bertin et al. 1994 with $-0.1< d<0$).

We assume that
 \begin{equation}
  \sigma^2({\mathcal A})\simeq\int_{\mathcal{A}} 2 \pi r dr \sigma^2(r) DV(r),
 \end{equation}
 where $r$ is the projected angle variable, $\mathcal{A}$ is the
projection on the focal plane of the solid angle subtended by the
instrument and $DV(r)$ is the $r^{1/4}$ law, appropriately normalised.
Under these assumptions
 \begin{equation}
  \sigma^2=\sigma^2_{\tx{ghff}} \frac{\int_0^{1/4} dx x
   \sigma^2(x r_{\tx{e}})DV(x r_{\tx{e}})}{\int_{\mathcal{A}} dr r \sigma^2(r)DV(r)}
   \equiv \sigma^2_{\tx{ghff}} {\mathcal B}^2(d).
 \end{equation}
The correcting factors were computed by taking into account the number
of pixels used during each spectrum extraction and are summarised in
Table~\ref{tab:scorr}.  Given the lack of information about the
 kinematic profiles, the correcting factors is taken to be
${\mathcal B}\equiv[{\mathcal B}(-0.1)+{\mathcal B}(0)]/2$, with an
error estimate $\delta {\mathcal B}\equiv [{\mathcal
B}(-0.1)-{\mathcal B}(0)]/2\sqrt{3}$ ($\approx$5\%).

\begin{table}
\caption{Kinematic correcting factors $\mathcal{B}$ with standard
deviations $\delta \mathcal{B}$ (see text) and final central velocity
dispersion $\sigma$ in \kms.}
\label{tab:scorr}
\begin{tabular}{ccccc}
galaxy & \sgh & $\mathcal{B}$ & $\delta \mathcal{B}$  & $\sigma$ \\
\hline
A & 232 & 1.07 & 0.04 & 248 \\
D & 271 & 1.14 & 0.08 &	309 \\
E & 175 & 1.10 & 0.06 & 192 \\
F & 195 & 1.08 & 0.05 & 211 \\
G & 226 & 1.09 & 0.05 & 246 \\
I & 166 & 1.11 & 0.06 & 184 \\
\hline
\end{tabular}
\end{table}

\subsection{Effective radii}

The radii $R_{\tx{eV}}$
and $R_{\tx{eB}}$ are calculated for the central rest wavelength of each
filter by linear interpolation/extrapolation in wavelength between the
measured radii in the WFPC2 filters F606W and F814W at their observed
central wavelengths.  We take the central wavelengths of B, V,
F606W and F814W to be 4400, 5500, 5935, and 7921~\AA, respectively,
resulting in the following explicit formulae for the angular sizes:
 \begin{eqnarray}
 r_{\tx{e}{\sc B}}=&\big\{r_{\tx{e}{\sc F606W}}[7921-(1+z)4400]\nonumber\\
 &+r_{\tx{e}{\sc F814W}}[(1+z)4400-5935]\big\}/1986 \\
 r_{\tx{e}{\sc V}}=&\big\{r_{\tx{e}{\sc F606W}}[7921-(1+z)5500]\nonumber\\ 
 &+r_{\tx{e}{\sc F814W}}[(1+z)5500-5935]\big\}/1986 \, .
 \end{eqnarray}
The typical correction is of order of 3\%.

\section{Results}
\label{sec:res}

We now consider the astrophysical implications of our measurements in
terms of the changes in scaling laws as a function of redshift.  We
emphasize that our sample, although small (only six galaxies with a
reliable measurement of the velocity dispersion), is unique in that it
focuses on {\it field} galaxies.  Therefore, we will first consider
our sample by itself (Subsection~\ref{ssec:ourres}), and then proceed
to compare with {\it cluster} galaxy measurements from the literature
(Subsection~\ref{ssec:compres}).

\subsection {Results from our sample: Scaling laws for field galaxies at
intermediate redshifts}
\label {ssec:ourres}

The input data for the galaxies in our sample are presented in
Table~\ref{tab:scorr} for the central velocity dispersion and in
Table~\ref{tab:fres} for the photometry.  Absolute magnitudes and metric
sizes in Table~\ref{tab:fres} are computed using $H_0$=50 $h_{50}$ \kms
$\tx{Mpc}^{-1}$, $\Omega=1$, $\Omega_{\Lambda}=0$.

\begin{table}
 \caption{Final corrected photometric values.  Effective radii are
expressed in $h_{50}^{-1}$kpc.  B and V refer to the corresponding Johnson
passbands. Quantities are computed using $\Omega=1$, $\Omega_{\Lambda}=0$.}
 \begin{tabular}{lcclccl}
galaxy&  $R_{\tx{eB}}$  &  SB$_{\tx{eB}}$   & $M_{\tx{B}}$   & $R_{\tx{eV}}$  & SB$_{\tx{eV}}$   &  $M_{\tx{V}}$ \\ 	
\hline
A  &     5.02   & 21.03	 & -21.23  & 5.73  & 20.37  &	-22.30	      \\
B  &     2.02   & 20.02	 & -20.00  & 1.94  & 18.95  &	-20.94	      \\
D  &     4.93   & 19.89	 & -22.16  & 4.22  & 18.65  &	-22.94	      \\
E  &     4.30   & 20.60	 & -21.13  & 4.60  & 19.83  &	-22.09	      \\
F  &     6.18   & 21.57	 & -20.97  & 6.76  & 20.85  &	-21.92	      \\
G  &     9.02   & 20.75	 &  -22.5  & 8.39  & 19.72  &	-23.29	      \\
I  &     3.52   & 19.89	 & -21.35  & 3.40  & 18.93  &	-22.20	      \\
O  &     5.61   & 20.70	 & -21.83  & 4.58  & 19.50  &	-22.61	      \\
P  &     3.58   & 20.29	 & -21.21  & 3.43  & 19.32  &	-22.09	      \\
\hline		  			      
\end{tabular}

\label{tab:fres}
\end{table}

We consider two scaling laws: the FP (Eq.~\ref{eq:FP}) and the
Kormendy relation \cite{K77}
\begin{equation}
SB_{\tx{e}} =a_1\log R_e+a_2.
\label{eq:HK}
\end{equation}  
The latter is useful for our purposes because it does not involve
internal kinematics, and thus it can be followed to higher redshifts
($ z = 0.647 $ for our sample), although it has a larger intrinsic scatter
than the FP and its coefficients are much more sensitive to sample
selection biases (see e.g., Capaccioli, Caon \& D'Onofrio 1992).

Because of the small number of points in our sample, we do not try to
rederive the parameters $ \alpha $ and $ \beta $ for the FP
independently, but we compare our sample of field ellipticals with the
local relations, adopting their slopes. In Figures~\ref{fig:allvscoma}
and \ref{fig:Ben} we compare our galaxies with a sample of Coma
ellipticals in Johnson V (Lucey at el. 1991, hereafter L91) and with a
recent determination of the FP in Johnson B, $\alpha=1.25, \beta=0.32,
\gamma=-8.895$, (B98).  The mean offset $\Delta \gamma $ (defined
as the mean of $\log R_e + \log h_{50} - \alpha \log \sigma-\beta {\tx{SB}}_{\tx{e}}
-\gamma$ over the data points) is 0.10 with respect to the L91 sample
(using the slopes given in L91, i.e. $\alpha=1.23$ and $\beta=0.328$)
and 0.07 in Johnson B.  We also compared our data {\it i)} to the 7
Samurai sample \cite{7S} obtaining $\Delta \gamma $=0.08 and 0.07,
respectively in Johnson B and V and {\it ii)} to the FP in Gunn $r$
(JFK96), finding $\Delta \gamma \approx 0.03$ with respect to the Coma
zero point, even if the colour corrections increase in this case the
uncertainty.
The rms scatter of our points is 0.16 in $\log R_e$ giving a standard
deviation of the mean value of about 0.07 in $\Delta
\gamma$. Considering other smaller sources of error such as the
peculiar velocity of Coma and the colour transformations we can
estimate the uncertainty on $\gamma$ to be $\approx 0.1$. The offset
found can be interpreted as a result of the evolution of the mean
$M/L$ ratio, $\Delta \log M/L\approx 0.1$. In other words, our field
galaxies are more luminous than a local elliptical with the same
effective radius and $\sigma$ (dimension and mass), consistently with
the expected evolution of stellar populations, and in agreement with
the results found by Bender et al. (1996), Ziegler \& Bender (1997),
Kelson et al.\ (1997), Schade, Barrientos \& L\'opez-Cruz (1997) and van
Dokkum et al.\ (1998), for the cluster environment.

If we exclude from the analysis the two galaxies {\bf E} and {\bf F}
with lower signal-to-noise spectra, the mean offset
becomes 0.15 with respect to the L91 sample, 0.11 to B98, 0.13 and
0.12 to the 7 Samurai B and V and 0.12 to JFK96. The scatter is
practically unchanged (0.16).  The change in the offset is not
negligible, but is not statistically significant. More data are needed
to extend this preliminary study to larger sample in order to overcome
small number fluctuations and to measure the offset and scatter with
smaller errors. Moreover when a larger sample will be available, it
will be possible to measure the offset in smaller redshift bins, thus
providing homogeneous subsamples. 

Clusters provide a homogeneous environment for the formation of
ellipticals. In contrast, there is no {\it a priori} reason to believe
that randomly selected field galaxies have formed at the same epoch,
and any differences in the formation redshift will be amplified with
look-back time. Therefore, one might have expected to find a larger
scatter for our field relation (if any relation was to be found at
all) than in the cluster FP. The scatter we find for our data
points (0.16 in $\log R_e$, using the local slopes) is in fact larger than
the scatter found by, e.g., L91 (0.075). Part of this scatter can be
explained because we are considering a small sample of galaxies at
different redshifts (and therefore different $\gamma$; from a
geometric point of view, the plane is thicker because is the
superimposition of parallel planes).  It is not clear however whether
there is a residual scatter, and a larger sample is required to
investigate the intrinsic variations of the formation and evolutionary
history for galaxies not in rich clusters.

\subsection {Comparison with other results: The evolution of the
Fundamental Plane}
 \label{ssec:compres}

Obviously, a larger sample of field galaxies is necessary to assess
whether they truly obey an FP-like relation and whether, and how much,
their properties vary with redshift.  However, a significant body of
data is now available on ellipticals at intermediate redshifts,
mostly from clusters, and it is instructive to look at our sample 
in the context of the cluster data as well.

\begin{figure}
\mbox{\epsfysize=8cm \epsfbox{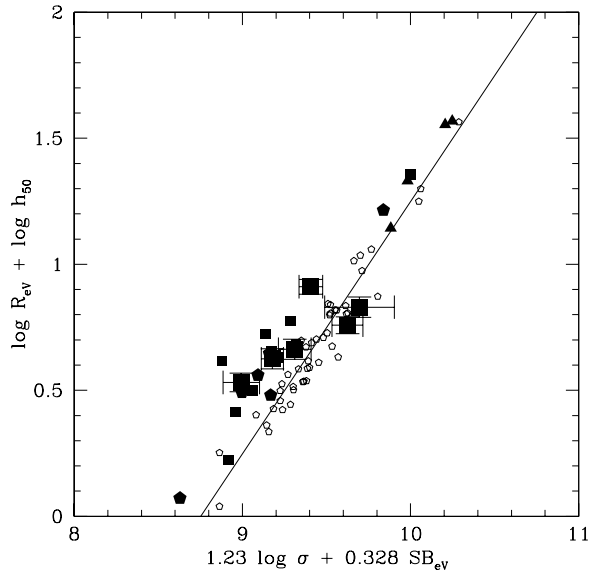}} 
\caption {The intermediate-redshift (filled) points plotted against
 the L91 (Lucey et al.\ 1991; empty pentagons) Coma cluster
 sample.  The filled points are from this paper (large squares), from
 van Dokkum and Franx 1996 (triangles), and from Kelson et
 al.~1997 (small squares and pentagons).  All surface
 brightnesses are in Johnson V. The line corresponds to the FP relation
 of L91.  The rms scatter in $ \log R_e $ is 0.14. The mean offset of
 the field ellipticals is 0.10 $ \log R_e $, of the cluster sample is 0.17 }
\label{fig:allvscoma}
\end{figure}

\begin{figure}
\mbox{\epsfysize=8cm \epsfbox{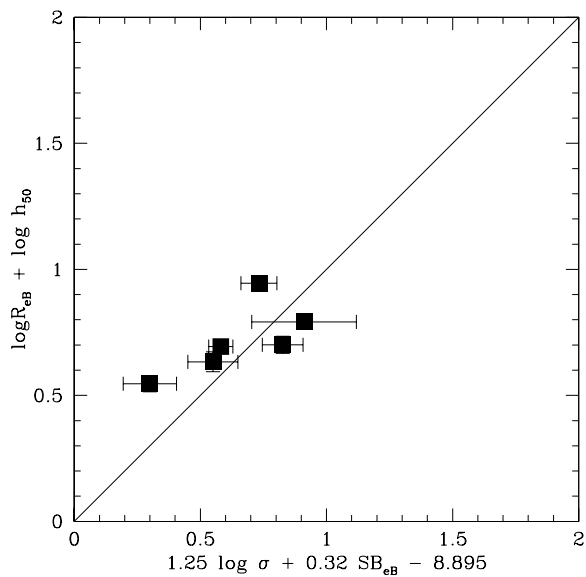}}
\caption {The field intermediate-redshift galaxies parameters in
Johnson B are shown as filled squares. The solid line is the FP
relation given by Bender et al. 1998. The mean offset is 0.07 in $\log
\tx{R}_{\tx{e}}$. The rms scatter in $\log \tx{R}_{\tx{e}}$ is 0.15.}
\label {fig:Ben}
\end{figure}

We consider here a total of 25 ellipticals from the papers of van
Dokkum and Franx (1996, 4 objects), Kelson et al.~(1997, 15 objects),
and from our sample (6 objects). 
 The first consideration is that these galaxies, much like
ours, fit reasonably well with the zero-redshift relations (see
Figure~\ref{fig:allvscoma}).  The zero point determined from all these
galaxies using the L91 slopes is $ -8.60 $
(Figure~\ref{fig:allvscoma}), very similar to the value obtained with
our six field galaxies alone.

\begin{figure}
\mbox{\epsfysize=8cm \epsfbox{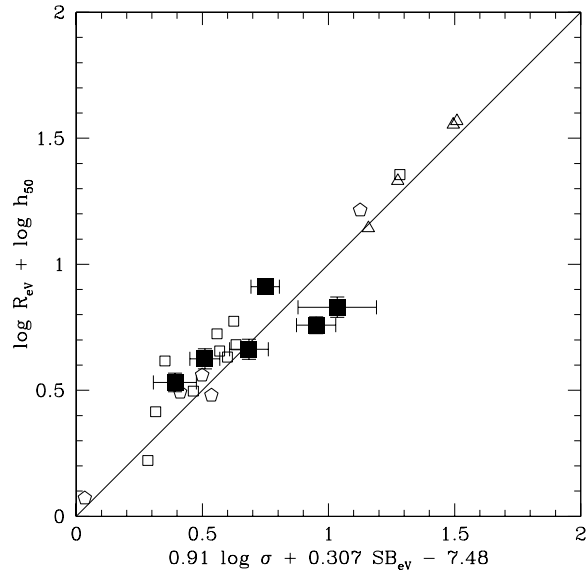}}
\caption {The intermediate-redshift points plotted against their own
best-fitting FP relation. The points are from this paper (large
squares), from van Dokkum and Franx 1996 (open triangles), and from Kelson
et al.~1997 (open squares and pentagons).  The rms scatter in $ \log
R_e $ is 0.10.}
 \label {fig:allvshigh}
\end{figure}

With this sample, it is also possible to determine the FP slopes $
\alpha $ and $ \beta $ independently of the low-redshift measurements.
The results depend on the method used for the minimisation ($\alpha$
especially, while $\beta$ is better determined). Using the technique
described in JFK96, which equally weights the three FP parameters, we
obtain $ \alpha = 0.91 $ and $ \beta = 0.307 $
(Figure~\ref{fig:allvshigh}). We estimate the errors to be
$\sigma_{\alpha}\approx 0.15$ and $\sigma_{\beta}\approx0.02$.  This
choice of the slopes reduces the scatter in $ \log~R_e $ from 0.13
(with respect to the local relation) to about 0.10.  The relation
$\alpha-10\beta+2=0$ (see Section \ref{sec:intro}) still holds within
the errors. If the reasoning is carried out with the assumption of the
single parameter $\eta$ ($L\propto M^{\eta}$), this would lead to
$\Delta \eta \approx -0.2$.  If confirmed with better statistics, this
variation might imply a differential passive evolution as a function
of mass, i.e. a mass-dependent formation redshift.

\begin{figure}
\mbox{\epsfysize=8cm \epsfbox{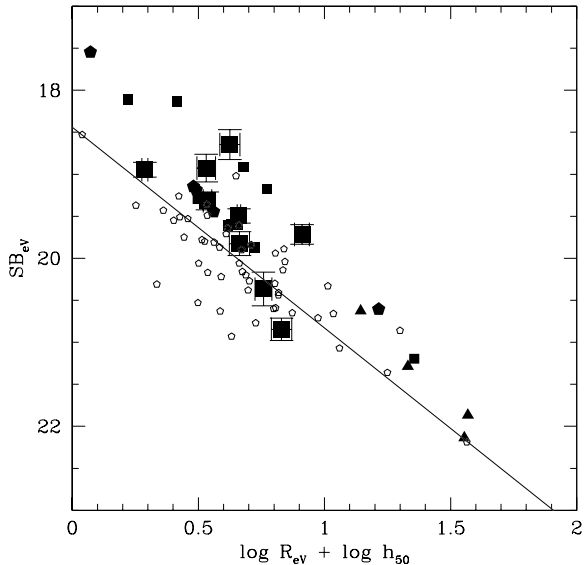}}
\caption{Surface brightness (SB$_{\tx{eV}}$) vs.  effective radius
(R$_{\tx{e}}$) for the same sample as in Figure~\ref{fig:allvscoma}
(filled symbols); the large squares are our data, the small squares
and the pentagons are the galaxies from CL1358+62 and MS2053-04
(Kelson et al.\ 1997), and the triangles are the four ellipticals of
CL0024 (van Dokkum \& Franx 1996).  The data of a sample of Coma
ellipticals (Lucey et al. 1991) are shown as empty pentagons for
comparison, and their best-fit Kormendy relation is overplotted. The
mean average offset is $\approx-0.5$ magnitudes.  Three galaxies are
also shown ({\bf B}, {\bf O}, {\bf P}) for which we have measured
photometry and redshift but no central velocity dispersion. }
\label{fig:totkor}
\end{figure}

In Figure~\ref{fig:totkor} we show the Kormendy relation between
SB$_{\tx{e}}$ and $R_{\tx{e}}$ for nine galaxies in our field sample (up to z=0.647)
plus the 25 other intermediate redshift cluster ellipticals.  The
intermediate redshift data are compared in the figure with the L91
sample of Coma ellipticals.  An average offset of $-0.52$ magnitudes is
found between the intermediate redshift cluster and field sample and the local
Coma sample.  The intermediate redshift galaxies are therefore on
average brighter (at the same $R_{\tx{e}}$), consistently with what is found using
the FP.  In order to determine the slope $a_1$, a sample with well-controlled
selection biases is required. The
intermediate redshift sample was not chosen to this aim and therefore
we do not attempt to derive the value of $a_1$.

More data and carefully selected subsamples are needed to assess if
the field Kormendy relation at intermediate redshift differs
significantly from the cluster one.

\section{Summary}

\label{sec:conc}

With the instrumental setup and the procedure described Sections 3, 4
and 5, we have measured the Fundamental Plane parameters of
intermediate redshift field elliptical galaxies.  As an important part
of the data-reduction procedure we have carried out accurate
K-corrections and spectroscopic aperture corrections.  Given the
importance of the template spectrum choice for the kinematic
fit, we have decided to make use of a wide set of spectra, checking
how the results vary with the spectral type, as described in Section
4, obtaining at the same time an estimate of the systematic error
produced by the choice of the template spectrum.

While the number of galaxies studied here is too small to draw firm
conclusions on the properties of field ellipticals at intermediate
redshift, the preliminary evidence suggests that: 

\begin{enumerate}
\item Our six field ellipticals at redshift $z\sim 0.3$,
are in agreement with the local FP relation, with a variation of
the zero point ($\Delta \gamma \approx 0.1$), and a scatter of 0.16 in
$\log R_{\tx{e}}$. This means that the stellar populations of our sample of field 
galaxies are brighter than the local ellipticals, with the same size and mass.
This data fit into a scenario in which our galaxies, at a look-back time 
of $\approx 4$--$5$ Gyrs, are evolving passively. The small sample and the
sample selection criteria do not allow any more general conclusion.

\item The FP obtained from our data and the cluster ellipticals at
intermediate redshift \cite{DF96,KDFIF} is well defined with a scatter
of 0.13 in $\log R_e$ with respect to the local relation. The full
intermediate redshift sample is large enough to perform the fit the
FP coefficients independently. Using the fitting technique used by JFK96,
we find $\alpha=0.91$ and $\beta=0.307$, and the scatter is reduced
to 0.10.  By interpreting the Fundamental Plane as a result of
homology, the virial theorem and the existence of a relation $L\propto
M^{\eta}$ \cite{3M}, different slopes at different redshift imply a
variation of $\eta$ with time, i.e., a mass-dependent evolution.  The
good agreement between the field and cluster galaxies suggests that
there are no significant differences in the history of the two
environments. More data are needed to perform separate fit to the two
subsamples.

\item The Kormendy relation between SB$_{\tx{e}}$ and $R_{\tx{e}}$ for
all the intermediate redshift data shows an offset of $\approx-0.5$
magnitudes in surface brightness with respect to the Coma ellipticals
(L91), in agreement with what is found using the FP. Even though
caution is needed, due to possible sample selection biases, the
variation of the Kormendy relation suggests that the evolution
found via the Fundamental Plane, extends to galaxies at higher
redshift. Carefully  selected samples of field
and cluster intermediate redshift are needed to study the
the Kormendy relation in greater detail and to
investigate the possible differences between the two environments.

\end{enumerate}
Further
discussion on the evolution of elliptical galaxies will be given in a
follow-up paper (Paper II; Stiavelli et al.\ 1999) with the help of
information on the line strengths. 

\section{Acknowledgments}

Tommaso Treu's work at Space Telescope Science Institute (STScI) was
financially supported by the STScI Summer and Graduate Student
Programs, the Scuola Normale Superiore (Pisa), the Italian Space
Agency (ASI), and by STScI DDRF grant 82216.  We thank an anonymous
referee for several valuable comments which significantly improved the
presentation of the kinematic measurement. The use of Gauss-Hermite
Fourier Fitting Software developed by R.~P. van der Marel and M. Franx
is gratefully acknowledged.

\end{document}